\documentclass[]{aa}

\usepackage{amsthm}
\usepackage{graphicx}
\usepackage{epsf,color}
\usepackage{natbib}
\usepackage{txfonts}
\definecolor{Mygreen}{rgb}{0.00, 0.72, 0.0}
\definecolor{Mypink}{rgb}{1.0, 0.0, 0.5}
\usepackage[breaklinks, citecolor=blue, linkcolor=blue, urlcolor=Mypink, colorlinks=true]{hyperref}
\usepackage{tablefootnote}
\usepackage{longtable}
\usepackage{subcaption}
\usepackage{graphicx}
\usepackage{rotating} % for sidewaysfigure environment
\usepackage{multirow} %mft
\usepackage{threeparttable}
\usepackage{xcolor}
\usepackage{soul}

\bibpunct{(}{)}{;}{a}{}{,}

\def\apjl{ApJ}%
\def\apjs{ApJS}%
\def\aap{A\&A}%

\definecolor{orange}{RGB}{255, 151, 112}

\begin{document}

\title{QUIJOTE-TFGI polarization calibration\\ \begin{large} Ground characterization and on-sky validation with Tau A and the Moon \end{large}}

\author{
Alessandro~Fasano \inst{\ref{iac},\ref{ull}}\thanks{Corresponding author: Alessandro Fasano, \url{alessandro.fasano@iac.es}}, 
Mateo~Fernández-Torreiro\inst{\ref{iac},\ref{ull}}, 
Guillermo~Pascual-Cisneros\inst{\ref{IPMU}, \ref{santander}},
Roger John~Hoyland \inst{\ref{iac},\ref{ull}},
Francisco Javier~Casas-Reinares\inst{\ref{santander}},
Ricardo Tanausú~Génova-Santos \inst{\ref{iac},\ref{ull}},
Michael William Peel\inst{\ref{iac},\ref{ull}},
Rafael~Rebolo-López \inst{\ref{iac},\ref{ull}},
José Alberto~Rubiño-Martín\inst{\ref{iac},\ref{ull}}}

\institute{
Instituto de Astrofísica de Canarias, E-38200 La Laguna, Tenerife, Spain\label{iac}
\and
Departamento de Astrofísica, Universidad de La Laguna (ULL), E-38206 La Laguna, Tenerife, Spain \label{ull}
\and
Instituto de Física de Cantabria (IFCA), Avda. Los Castros s/n, 39005 Santander, Spain \label{santander}
\and
Kavli Institute for the Physics and Mathematics of the Universe (WPI), The University of Tokyo Institutes for Advanced Study, The University of Tokyo, Kashiwa, Chiba 277-8583, Japan \label{IPMU}
}  

\abstract
{Polarization measurements in the 30--40 GHz range are essential to improve the understanding of Galactic synchrotron radiation and potential anomalous microwave emissions. Ultimately, this advancement will aid in detecting primordial B modes. To accomplish this, it is crucial to achieve unprecedented performance in instrumentation. Specifically, accurately calibrating instrument responses and polarization angle is an essential step in state-of-the-art setups. The modeling and mitigation of systematic effects are critical in this context.}
{Our objective is to characterize the QUIJOTE Thirty and Forty GHz instrument (TFGI), calibrate it with a reference calibration signal on the ground, compare our results with on-sky calibration based on bright sources, and study the stability of the calibration parameters over time.}
{First, from the ground, we fit the data using a reference calibration signal (a diode) introduced to resolve degeneracies among the various instrument angles. Finally, we utilize on-sky observations of Tau A and the Moon to validate the results.}
{By creating calibration datasets obtained with the reference diode, we evaluate the data quality and quantify phase switch errors to account for the fine polarization response. We also utilize Tau A and Moon observations to calibrate the system's response and stability over time. 
In addition, we calculate the refraction index of the Moon to be $n_{\rm{Moon}} = 1.209 \pm 0.007\, \text{(stat)}\,\pm0.009\,\text{(sys)}$ at 31\,GHz under smooth-surface assumption.}
{The results from fitting the instrument phase-switch error angle align with 0\,deg at 2$\sigma$ precision, indicating that no further correction is required within a few percent precision. The calibrations with astrophysical sources (Tau A and the Moon) yield consistent results that constrain the polarization angle and responsivity.
The polarization efficiency aligns well with ground measurements and the Tau A characterization, whereas the Moon-based calibration is more affected by systematics.
We find hints of responsivity variations over time, although the relative responsivity between channels is found to remain stable.
In the future, we conclude that installing a live calibrator will enhance performance by continuously monitoring responsivity and, in turn, improving the mitigation of systematic effects.}


\titlerunning{ QUIJOTE-TFGI calibration }
\authorrunning{Alessandro~Fasano et al.} 

\maketitle

\section{Introduction}
\label{sec:introduction}

Studying the cosmic microwave background (CMB) is essential for constraining the fundamental parameters that shape the evolution of the Universe.
By meticulously mapping the anisotropies in the CMB, cosmology has gained invaluable insights into the underlying cosmological principles, thereby confirming the Lambda cold dark matter ($\Lambda$CDM) paradigm~\citep[e.g.,][]{1996ApJ...460....1K,2013ApJS..208...20B,2013ApJS..208...19H,2020A&A...641A...6P}. Notably, the Planck ESA satellite has significantly contributed to this context, producing full-sky maps with unparalleled resolution and sensitivity (\citealp{2016A&A...594A...6P}).

With the completion of these groundbreaking observations, attention has shifted towards investigating the polarization of the CMB (\citealp{2016arXiv161002743A,2019JCAP...02..056A,2019JLTP..194..443H}). Despite its signal being orders of magnitude fainter than total intensity, advancements in instrumentation now permit the study of polar signals with unprecedented sensitivity. The polarization pattern of the CMB, typically decomposed into E-modes and B-modes (\citealp{1997PhRvL..78.2054S,1997PhRvD..55.7368K}), carries critical information about the early Universe's conditions, including the presence of primordial gravitational waves generated during inflation.%(\citealp{2020PhRvL.125v1301M,2020A&A...634A.100A,2022JCAP...04..029V}). % these are for polarization angle
While satellite-based experiments are required to measure the largest angular scales, ground-based observations have also made significant contributions to the field, especially at small angular scales where satellite telescopes struggle to achieve the necessary angular resolution. The international thrust has deployed in the last decade, among others, the Atacama B-mode Search experiment (ABS; \citealp{simon2016characterizing, kusaka2018results}), the Atacama Cosmology Telescope (ACT; \citealt{li2021situ,qu2024atacama}), the BICEP/\textit{Keck} \textit{Array} program \citep{hui2018bicep, BK-XV20}, the Cosmology Large Angular Scale Surveyor (CLASS; \citealt{essinger-hileman14spie,eimer23}), the GroundBIRD telescope (\citealp{2024SPIE13102E..05T}), the {\scshape Polarbear} experiment (\citealp{inoue2016polarbear, adachi2022improved}, the predecessor of the Simons Array experiment; \citealt{stebor2016simons}), the South Pole Telescope (SPT; \citealt{chown2018maps, sobrin2022design}), {\scshape Spider} (\citealp{2022ApJ...927..174A,2024SPIE13102E..04S}), and the Q \& U Bolometric Interferometer for Cosmology (QUBIC; \citealp{2024EPJWC.29300030M}).
%and Taurus (\citealp{2024SPIE13094E..32M}).

In this context, the Q-U-I JOint TEnerife (QUIJOTE) CMB experiment aims to achieve two primary scientific objectives: firstly, 
to measure the tensor-to-scalar ratio (r) if larger than $\sim$0.05 (\citealp{2017hsa9.conf...99R}); and secondly, to characterize polarized foregrounds (\citealp{2009A&A...503..691B}), particularly at low frequencies, shedding light on the synchrotron and anomalous microwave emissions.
The current best upper limit on $r$ is $r < 0.032$, at a 95\% confidence level (\citealp{2022PhRvD.105h3524T}). Current-generation experiments like the Simons Observatory small aperture telescopes (SO-SATs) aim at constraining $r$ with $\sigma(r) = 0.003$ or better (\citealp{2019JCAP...02..056A}), while next-generation telescopes like LiteBIRD will attempt a $r < 0.002$, at a 95\% confidence level  \citep{2023PTEP.2023d2F01L}.

Ensuring the precise calibration of the polarization angle presents a significant challenge for upcoming CMB polarization experiments.
Any unaccounted systematic effects may lead to spurious signals exceeding the faint CMB B-mode signal.
The precision of angle calibration holds particular significance here, as any associated errors may lead to contamination, converting the stronger E-modes into the B-modes signal.
The international community is currently engaged in a thrust to develop artificial calibrators for the next-generation CMB experiments (\citealp{Cas22,2023RScI...94l4502M,2024EPJWC.29300044R,2024SPIE13102E..24C}) to improve the absolute polarization angle calibration to the suitable precision, better than 0.1\,deg. For instance, an experiment targeting $\rm{r}\sim10^{-3}$ requires a precision better than 0.2\,deg (\citealp{2020A&A...634A.100A}).
The astrophysical calibrators are not yet calibrated at this precision, with the principal candidate being Tau A (also known as the Crab nebula), which presents a polarization angle of $-88.26\pm0.27$\,deg (\citealp{2020A&A...634A.100A,2022JCAP...04..029V}).

In addition to polarization angle calibration, an equally crucial aspect is the precise calibration of detector responsivity. Any uncertainty in the responsivity directly affects the amplitude of the measured Stokes parameters, thereby introducing a multiplicative bias in the polarization power spectra. For next-generation CMB experiments aiming to detect primordial B-modes at levels of $r<10^{-3}$, the required responsivity calibration must be at the sub-percent level, typically better than 0.5\%, to avoid systematic errors that could bias the signal (\citealp{2021JCAP...05..032A}). Responsivity mismatches across detectors can also induce leakage from temperature anisotropies into polarization.
Precise calibration between frequency channels is also required to ensure a reliable component separation and then minimize contamination on the measurement of the cosmological signal from foreground residuals \citep{2025JCAP...01..019C}. Finally, differential responsivity errors distort the reconstructed Stokes parameters by introducing mixing between $I$, $Q$, and $U$, which in turn leads to leakage between $T$, $E,$ and $B$ \citep{2007MNRAS.376.1767O}.

In this work, we describe the calibration of the TFGI system by characterizing its central detector, Pixel 23 at 31\,GHz. This detector is the one located in the center of the focal plane, and we use it as a benchmark for the full focal-plane calibration.
This choice is driven by the central pixel’s optimal optical performance; however, the results are not assumed to be directly representative of off-axis detectors. A full focal-plane calibration is beyond the scope of this work, which primarily aims to establish a calibration methodology for TFGI.
In Sect.~\ref{sec:instrument}, we provide an overview of the TFGI instrument. Section~\ref{sec:ground_calibration} details the ground measurement performed with the instrument mounted on the telescope but without sky scanning, including its specific setup and mathematical framework (Sect.~\ref{sec:cal_setup_lab}), data structure (Sect.~\ref{sec:cal_lab}) for polarization efficiency (Sect.~\ref{sec:pol_eff_lab}), responsivity (Sect.~\ref{sec:gain_calibration_lab}), polarization angle (Sect.~\ref{sec:pol_ang_lab}), and phase error (Sect.~\ref{sec:phase_error_calibration}) characterization.
Section~\ref{sec:on_sky_calibration} presents the results from an on-sky calibration and response stability using Tau A (Sect.~\ref{sec:cal_TauA}) and the Moon (Sect.~\ref{sec:cal_moon}).
Section~\ref{sec:comparison} provides a comparison of the results across different setups.
In addition, we calculate the refraction index of the Moon in Appendix~\ref{sec:moon_refraction_index}.
Finally, the conclusions are presented in Sect.~\ref{sec:conclusions}.

\section{The QUIJOTE-TFGI instrument}
\label{sec:instrument}

The heart of the QUIJOTE project lies in its instrumental capabilities, notably embodied in the Thirty and Forty GHz Instrument (TFGI). We refer to the two instruments (TGI and FGI) as TFGI in this work from now on.
This instrument combines in the same cryostat the detectors of the Thirty-GHz Instrument (TGI), designed to operate within the frequency range of 26--36\,GHz \citep{TGI_SPIE2014, TGIcontrolsystem, TGIcryomechanics, TFGIreceivers_artal}; and the Forty-GHz Instrument (FGI), operating at 35--47\,GHz \citep{status2016SPIE, Rubino17, TFGIreceivers_artal}.
The TFGI has better angular resolutions than Planck at the same frequencies by a factor of $\sim$1.6 (\citealp{2016A&A...594A...4P}): 31\,GHz TGI detectors have full width at half maximum (FWHM) values of 21\,arcminutes, while 41\,GHz FGI detectors go down to 17\,arcminutes.

\begin{figure*}[ht]
\centering
\includegraphics[width=.99\textwidth]{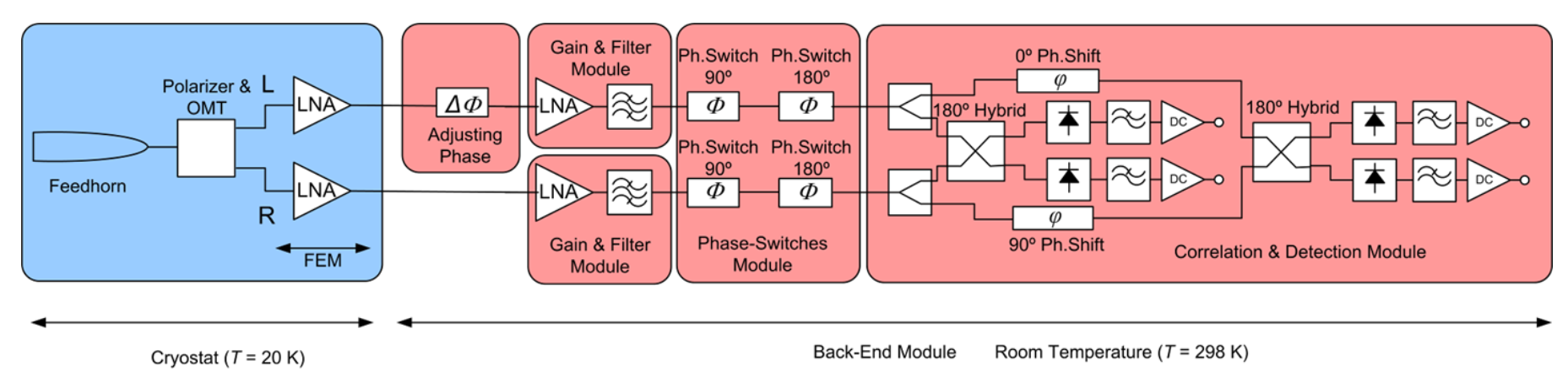}\\
\includegraphics[width=.99\textwidth]{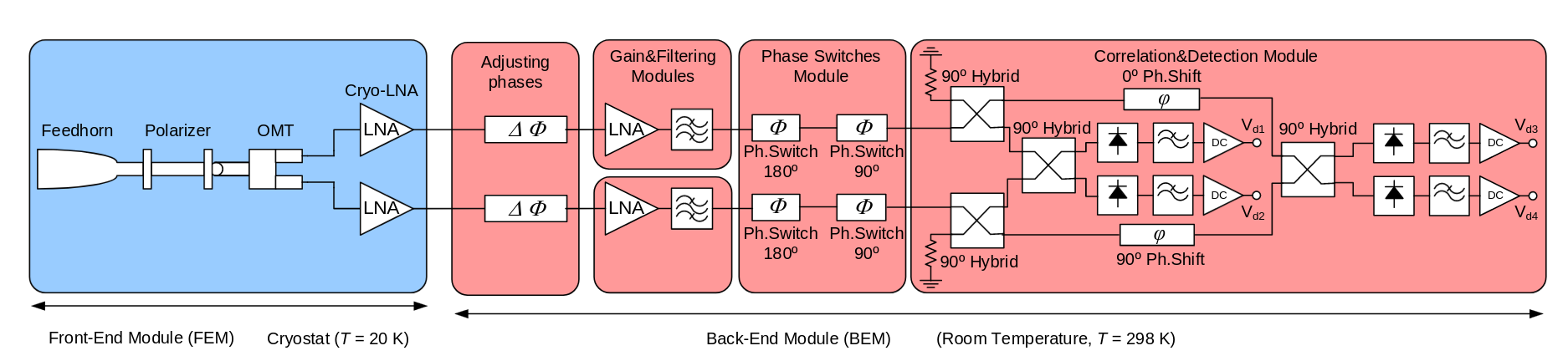}
\caption{
Block diagrams of the receivers (one detector). Up: Thirty GHz Instrument. Bottom: Forty GHz Instrument. The diagram is taken from \cite{TFGIreceivers_artal}.
}
\label{fig:tfgi_schematic}
\end{figure*}

Its intricate design and modular construction enable precise measurements of CMB polarized signals while minimizing systematic errors associated with conventional radiometer setups. The TFGI receivers are polarimeters capable of simultaneously measuring the Stokes parameters ($I$, $Q$, and $U$). 
%JARM: Text removed. Although the instrument is sensitive to the $V$ Stokes parameter, $V$ is not considered because the CMB is not expected to exhibit circular polarization.
% AF: I let $V$ in texit, \rm{V} is used afterward as output in volts
The block diagrams in Fig.~\ref{fig:tfgi_schematic} show the configuration of each receiver.

The TFGI cryostat can host up to 29 detectors, which may be all TGI, all FGI, or a combination of both.
%The full system will consist of 29 detectors in TGI and 29 detectors in FGI.
In this study, we focus on Pixel 23 of TGI, which, in the current configuration, is located at the center of the focal plane and exhibits the best optical performance.
The TFGI comprises interconnected modules, each meticulously engineered to fulfill specific functions within the receiver system. 
Starting with the cryogenic front end, cooled to temperatures between 10 and 20\,K, this module houses critical components such as feedhorns, polarizers, ortho-mode transducers (OMTs), and low-noise amplifiers (LNAs). These components establish the foundation for efficient signal reception and amplification, ensuring minimal loss and maximal sensitivity. Their wide-band design (with a relative bandwidth of $\sim$30\%) and near-theoretically ideal characteristics provide the most balanced and efficient polar extraction.
Continuing through the radio-frequency gain stage, which incorporates bandpass filters and amplifiers for both polar branches, the TFGI maintains fidelity in signal branching while enhancing signal strength. Subsequent modules contain essential components, such as broadband phase switches and complex correlators, that facilitate precise measurements of Stokes parameters through signal processing.
A detailed description of the detectors is provided in \cite{TFGIreceivers_artal}. 

The phase-switch module implements phase modulation in two identical branches, each containing a 180\,deg and a 90\,deg broadband phase switch. Each branch comprises two outputs, referred to as channels. Therefore, each detector presents four channels. 
Signal correlation essentially involves the addition and subtraction of broadband microwave signals by a system based on 90\,deg or 180\,deg hybrid couplers and a fixed 90\,deg phase shifter in one of the branches.
According to the phase-switch state, out of the four possible states (0, 90, 180, and 270\,deg), the detected signals from the correlation-detection module (denoted as voltage outputs) are proportional to different combinations of Stokes parameters.

For each nominal phase state, there are four distinct combinations of the four switches that result in the same effective phase. Each of these is recorded individually in the time domain and is referred to as an engineering state. Operationally, the four engineering states corresponding to a given phase state are averaged to produce a single value, known as the scientific mode, which is used in subsequent analysis. 

In summary, each detector comprises four channels, each recording a signal -- referred to as an output -- corresponding to a sequence of four phase states, known as science states. Each science state is computed by averaging four equivalent engineering states.
The data are synchronously sampled in parallel across the four channels at a rate of 160\,kHz, with 40 samples acquired for each engineering state. This results in a complete cycle of 16 consecutive engineering states being recorded every 4\,ms. We employ the science state signal directly in this work.

TFGI was first assembled with 29 receivers in 2016 and achieved its technical first light in May 2016 \citep{status2016SPIE, Rubino17}, but observations were interrupted due to cryostat leak issues.
After preliminary commissioning phases carried out between 2018 and 2019, the TFGI was finally installed at the second QUIJOTE telescope (QT2) and underwent its main commissioning phase from November 2021 to October 2022 with only 7 pixels, 4 at 31\,GHz and 3 at 41\,GHz. 
Scientific results from this commissioning phase will be presented in a separate paper (\citealp{tfgi_commissioning}). The data analyzed in this work correspond exclusively to this latter commissioning period.

\section{Ground calibration}
\label{sec:ground_calibration}

\subsection{Setup}
\label{sec:cal_setup_lab}

\begin{figure}[ht]
       \centering
        \includegraphics[width=.28\textwidth]{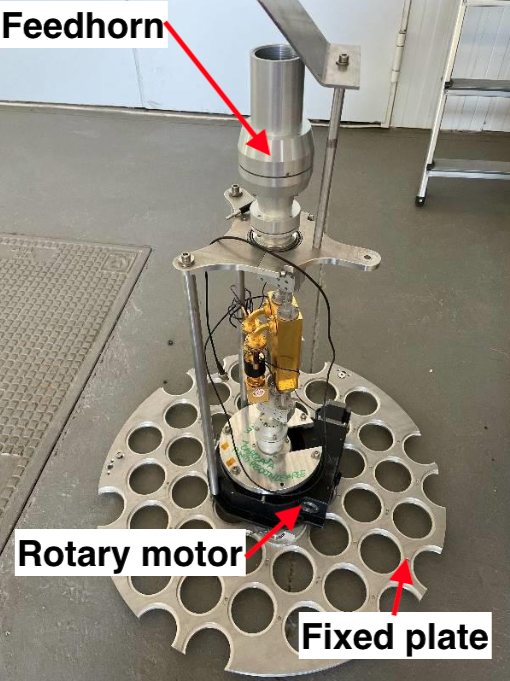}
        \caption{Photo of the MFCI mounted on the calibration alignment plate. The diode used introduces the known signal into the TFGI detectors. The fixed plate is placed on top of the focal plane and defines the different detector locations where the diode can be positioned. The rotary motor allows the modification of the calibration diode's orientation ($\gamma$) in 22.5\,deg steps to study different linear polarization inputs. A detailed description is presented in \cite{mateo_phd}.}
        \label{fig:calibrator}
\end{figure}

The calibration process involves positioning the so-called multi-frequency calibration instrument (MFCI) in front of the detector and acquiring calibration data, as illustrated in Fig.~\ref{fig:calibrator}. The MFCI consists of two feedhorns and OMTs, similar to the cryogenically cooled feedhorns and OMTs of the TFGI, and is connected back-to-back through two identical -20\,dB waveguide couplers. The calibration signal is introduced through one coupler, while the other is terminated with a room-temperature load. The horns are aligned through a concentric stepper motor axial drive, and this setup is designed to be attached to a detector-centered polarization grid known as a fixed plate. Additionally, a tilted aluminum flat mirror points the horn arrangement toward the sky when the cryostat is mounted on the focal plane of the QT2. In this way, the background radiation during a calibration run comes from the sky.
The injected signal corresponds to the diode’s equivalent temperature when ON with a 20\,dB coupling ratio.
A wideband noise diode (Keysight 346CK01\footnote{\url{https://www.keysight.com/us/en/product/346CK01/noise-source-1-ghz-50-ghz.html}}) was selected for its coverage of both FGI and TGI bands.

For installation, the MFCI is preassembled horizontally using a digital inclinometer and mounted onto the telescope focal plane with 0.2\,deg mechanical repeatability of the MFCI mounting angle when reinstalling the calibration plate. A dedicated acquisition system controls the stepper motor, the diode, and the phase switches. Polarization angles are referenced with respect to the diode’s E-field and defined as positive for anticlockwise rotation from the sky toward the instrument.
The alignment plate is secured over the cryostat focal plane, and the MFCI is attached with spring-loaded clamps, ensuring the mirror maintains its skyward orientation. Electrical connections for the motor, diode, and switches are routed through dedicated cabling.

The 16 possible phase switchers' states could present a misalignment with respect to the desired angle (0, 90, 180, and 270\,deg). Due to their non-idealities, deviations from the expected polar measurement occur. However, due to the complete coverage of all phase errors, any systematic error is generally canceled when all states are summed with the correct error for a given Stokes parameter. Notwithstanding, we want to prove this and characterize the precision of such an instrumental defect, in Sect.~\ref{sec:phase_error_calibration}.

\subsubsection{Process of polarization characterization and mathematical framework}
\label{subsec:eq}

Each phase switch is designed to introduce a phase shift as close as possible to its nominal value -- either 90\,deg or 180\,deg. However, imperfections in the hardware can lead to small deviations, denoted as $\epsilon$, from the nominal phase shift, and may also introduce frequency-dependent variations across the band. These deviations must be carefully characterized and, if necessary, corrected to ensure accurate signal reconstruction and minimize systematic errors.
In principle, averaging over the four equivalent engineering states to produce a science state (see Sect.~\ref{sec:instrument}) can help reduce the impact of phase errors, assuming that the errors are random. However, a residual systematic contribution may remain, particularly if the phase error exhibits some degree of coherence or bias across states.

To break the degeneracies between the phase error ($\epsilon$) and other instrument-induced angles, we built a dataset where the polarization angle of the calibration source is rotated by an angle $\gamma$, by a polarizer. 
This polarizer introduces an angle that allows us to repeat the same measurement with a controlled shift. The values for $\gamma$ range from -45.0 to 67.5\,deg in steps of 22.5\,deg, (-45.0, -22.5, 0.0, 22.5, 45, and 67.5\,deg). This approach yields six measurements for each phase state, thereby adding five equations to the system for each fixed phase state.
Mathematically, in a simplified form by assuming that the incoming light is fully polarized, we can express the Stokes-parameters equations as:
\begin{align}
P&=\sqrt{ Q^2 + U^2 } = I \label{eq:assumptions} \\
Q&=P \cdot \cos{ \left[ 2 ( \gamma - \phi_c ) \right] } \nonumber \\
U&=-P \cdot \sin{ \left[ 2 ( \gamma - \phi_c ) \right] } \nonumber
\end{align}
\noindent where $I$, $Q$, and $U$ represent the Stokes parameters, $\gamma$ is the polarization angle of the calibration source, and $\phi_c$ is the reference polarization angle of the TFGI polarimeter. 
The minus sign in $U$ is due to the instrument’s coordinate frame: the TFGI reference axis is mirrored relative to the standard sky-frame definition, which flips the sign of the measured $U$ component.
Introducing Eq.~\eqref{eq:assumptions} into the set of equations that define the TFGI channels' responses produces a new set of equations that relate the output of each channel to the input Stokes parameters. Consequently, the simplified (see \citealp{mateo_phd} for the extended formalism) TFGI equations -- in the assumption of 100\%-polarized signal -- for the TGI and FGI channels (output voltages of each detector or polarimeter), respectively (distinguished in the superscript), are:
\begin{subequations}\label{eq:new_equations}
\begin{align}
\rm{V}^{\rm{TGI,FGI}}_{1,4} &= \frac{1}{2} g_{1,4} I \left[ 1 - \eta\sin \left[ \delta + \epsilon_{1,4}(\delta) + 2 ( \gamma - \phi_{ \rm{ c;1,4 } } ) \right ] \right] \label{eq:1} \\
\rm{V}^{\rm{TGI,FGI}}_{2,3} &= \frac{1}{2}g_{2,3} I \left[ 1 + \eta\sin \left[ \delta + \epsilon_{2,3}(\delta) + 2 ( \gamma - \phi_{ \rm{ c;2,3 } } ) \right ] \right] \label{eq:2} \\
\rm{V}^{\rm{TGI,FGI}}_{3,1} &= \frac{1}{2} g_{3,1} I \left[ 1 + \eta\cos \left[ \delta + \epsilon_{3,1}(\delta) + 2 ( \gamma - \phi_{ \rm{ c;3,1 } } ) \right ] \right] \label{eq:3} \\
\rm{V}^{\rm{TGI,FGI}}_{4,2} &= \frac{1}{2} g_{4,2} I \left[ 1 - \eta\cos \left[ \delta + \epsilon_{4,2}(\delta) + 2 ( \gamma - \phi_{ \rm{ c;4,2 } } ) \right ] \right] \label{eq:4}
\end{align}
\end{subequations}
\noindent where V is the output in volts\footnote{Not to be mistaken this time with the Stokes $V$ parameter.}, $g$ is the gain, $\eta$ is the polarization efficiency, $\delta$ is the phase state of each polarization state (0, 90, 180, and 270\,deg), and $\epsilon$ is the correction angle that represents the deviation from the nominal phase value assumed constant over the band. The subscript indicates the channel, with a comma distinguishing between TGI and FGI.
Also $g$, $\epsilon$, and $\phi_{\rm{c}}$ differ between TGI and FGI.
Operationally, we use Eq.~\eqref{eq:3} for all channels in the on-sky analysis, since the four-channel equations are mathematically equivalent and differ only by successive 45-degree rotations; adopting a single convention therefore simplifies the on-sky data processing without loss of generality.

\subsection{Data}
\label{sec:cal_lab}

\begin{figure}[ht]
        \centering
        \includegraphics[width=.50\textwidth]{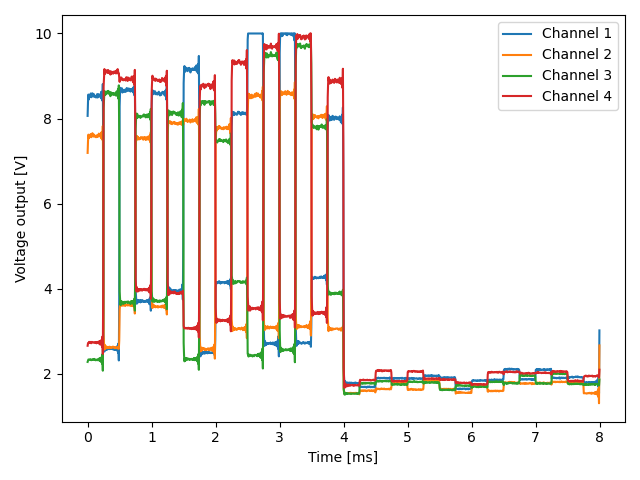}
        \caption{Example of a full data block, $\sim$8\,ms. The left part of the voltage output shows the diode ON, while the right part shows the diode OFF. Each color identifies a channel output. The signal exhibits saturation, reaching a value of 10\,V, in some cases.}
        \label{fig:voltage_diode_dark}
\end{figure}

As outlined in Sect.~\ref{sec:cal_setup_lab}, the calibration tests involve assessing the signal response while manipulating the angle of a polarizer placed between the feedhorns of the polarimeters and the calibration source. Initially, the diode source is activated (these data are referred to as diode ON). Subsequently, the source is deactivated (these data are referred to as diode OFF) to measure the background signal. Figure~\ref{fig:voltage_diode_dark} displays an example of a complete block of data ($\sim$8\,ms), where the first part presents the diode signal (diode ON) and the right part represents the pure background signal (diode OFF). The steps indicate the different 16 phase states, and the colors identify the channels. The outputs from the different channels vary at different levels due to the varying gains of the system, which we characterize in Sect.~\ref{sec:gain_calibration_lab}.

By averaging the blocks for each phase state, we compute the pure diode signal ($\rm{ V_{diode}}$) as:
\begin{equation}
    \rm{V}_{\rm{diode}} = V_{\rm{ON}} - V_{\rm{OFF}}
    \label{eq:diode}
\end{equation}

\noindent where V$_{\rm{ON}}$ is the diode ON signal (left side of Fig.~\ref{fig:voltage_diode_dark}) and V$_{\rm{OFF}}$ is the diode OFF signal (right side of Fig.~\ref{fig:voltage_diode_dark}).

First, we utilize a straightforward dataset for TGI where the external polarizer is not included (i.e., no $\gamma$ angle), named 
%Pixel23\_polcal1-22-04-06-11-57-59
Pixel23\_2022\_April\_06\_0. This more straightforward dataset is used to assess data quality (Sect.~\ref{sec:data_assessment}) concerning response saturation and stability over time. Finally, for the $\epsilon$ characterization, we concentrate on three specific datasets where we introduce the external polarizer (with $\gamma$ angles of -45.0, -22.5, 0.0, 22.5, 45, and 67.5\,deg): Pixel23\_2022\_April\_06\_1, Pixel23\_2022\_April\_06\_2, and Pixel23\_2022\_April\_07\_3. The visualization of the data structure is present in Appendix~\ref{app:data_structure}.

\subsubsection{Data assessment with diode reference}
\label{sec:data_assessment}
Data assessment is crucial for ensuring the reliability and accuracy of the scientific results derived from observations. This process involves carefully scrutinizing the raw data collected by the instrument, identifying potential artifacts or noise sources, and implementing corrective measures to reduce their impact on the final analysis.
\paragraph{Response saturation}
First, we evaluate the quality of the data by visually inspecting the time-ordered information. As expected, the signal reaches its maximum value at 10\,V. However, a non-linear response could occur before reaching this threshold, resulting in a loss of information on one side and introducing systematic effects in the response evaluation on the other.
We observe high signal levels in the diode ON reference data, indicating the presence of saturation-related systematics. In contrast, the signal in the diode OFF reference is significantly lower (see Fig.~\ref{fig:voltage_diode_dark}).
We do not modify or correct the signal in the presence of saturation in this work.

\paragraph{Stability in the time stream}
The next critical step involves assessing the stability of the data, which refers to the system's ability to consistently translate the same incoming signal into the same electric (readout) signal across different instances.
The total measurement duration spans 1\,minute. To evaluate stability, we resample each phase state (comprising 40 points each at a sampling rate of 160\,kHz) by calculating their averages. Subsequently, we plot the time series of these averages and apply linear regression to determine the slope. A slope close to zero indicates a stable output signal.

\begin{figure}[ht]
        \centering
        \includegraphics[width=.50\textwidth]{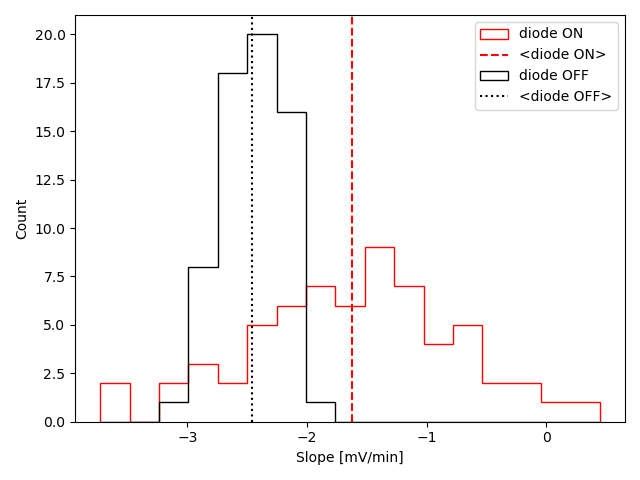}
        \caption{The histogram of the slopes that quantify the signal stability in a data acquisition of 1 minute. Each count is defined as a function of the slope obtained from linear fitting. The slope values span a range $\sim$$-2.5$--0.5\,mV/minute for the diode ON case, indicating a lack of significant stability issues within 1 minute of integration. }
        \label{fig:stability_histogram}
\end{figure}

We perform a statistical analysis of response stability. Across all channels and phase states, the signal is consistently reproduced throughout the 1-minute duration, showing no noticeable drift or instability. The diode ON data corresponds to the activation of the input calibration signal. In contrast, the diode OFF data are collected when the reference source is off, and the detector registers the background.
The results are shown in Fig.~\ref{fig:stability_histogram}.

The data present a linear regression slope of a few hundred mV/minute; $-1.6\pm0.8$\,mV/minute in the diode ON case and $-2.5\pm0.3$ mV/minute in the diode OFF case. The uncertainties are calculated as the standard deviation of the counts. This leads to a drift in the signal smaller than 2\,mV in a minute. Equivalent to $\sim$50\,mK by applying the average on-sky responsivity reported in Sect.~\ref{sec:comparison_responsivity}. This instability is likely due to a combination of gain drift and the instability of the optical load, originating from the background and the diode.
Although this small drift is detectable, it has a negligible impact on the results discussed here. 
Firstly, it is comparable to the white-noise level. Secondly, it is subdominant with respect to other systematic effects in the calibration. In particular, differences between the four phase-state configurations, which should nominally produce identical values, dominate the uncertainty budget at the few-percent level, whereas the measured drift over one minute is only at the millipercent level.

\subsection{Polarization efficiency calibration}
\label{sec:pol_eff_lab}
The polarization efficiency of a system quantifies how effectively the instrument produces, transmits, or manipulates polarized light. Polarization efficiency depends on several factors. First, high-quality polarizing elements block or transmit light based on its orientation. Accurate alignment of optical components within the system is crucial to prevent polarization distortion or loss. Additionally, the properties of the materials used in optical components affect polarization characteristics. Efficiency may vary across the electromagnetic spectrum, which can impact system performance. Environmental conditions, such as temperature and humidity, can influence polarizing elements and polarization efficiency. Finally, system design, including component arrangement and the choice of elements, is essential for optimizing efficiency and minimizing unwanted polarization effects. Quantifying polarization efficiency involves measuring the degree of polarization of light at different stages of the system and comparing it to the input (see \citealp{Tinbergen_1996} for a detailed description of light polarization). In the TFGI case, the frequency dependence of polarization efficiency on wavelength is not fully appreciated, as it is integrated across the bandpass. 
%We might further refine this calculation by measuring it in the future. 
Since the calibration signal is fully polarized, the polarization efficiency is expressed as
 $\eta = \sqrt{ Q^2 + U^2 } / \textit{I}$ where we can obtain the $I$ Stokes parameter from Eq.~\eqref{eq:new_equations} as
\begin{equation}
    I = \left[ \rm{V}(0\,\rm{deg}) + V(90\,\rm{deg}) + V(180\,\rm{deg}) + V(270\,\rm{deg}) \right] / 2 
    \label{eq:intensity}
\end{equation}
\noindent where the angle values in the brackets correspond to $\delta$, and the $Q$ and $U$ Stokes parameters are defined in Appendix~\ref{app:eq_pol} for all the sets of equations.
We utilized the voltage output of a pure diode signal ($\text{V}_{\rm{diode}}$), by calculating it with Eq.~\ref{eq:diode}.
We conduct statistical analyses for each channel and polarizing angle ($\gamma$), alongside the three selected datasets, to calculate the polarization efficiency in percentage ($\eta$). The average results are presented at the top of Table~\ref{tab:pol_angle_total}. The uncertainty is calculated through the standard deviation across different datasets; the detailed report is in Table~\ref{tab:all_pol_eff}.

The system exhibits a mean polarization efficiency across the channels of $\eta=95\pm2$\%, indicating a polarization loss\footnote{The polarization loss is defined as: $100\%-\eta$.} of under $\sim$10\%. The uncertainty of the mean polarization efficiency is determined as the standard deviation across the value per channel. The extensive results, presented by polarizing angle and dataset, are reported in Table~\ref{tab:all_pol_eff} of the Appendix~\ref{app:eq_pol}. These values are comparable to the Multi-Frequency Instrument (MFI) values (Table~6 in \citealp{mfiwidesurvey}).

\subsection{Responsivity calibration}
\label{sec:gain_calibration_lab}

We utilize the dataset Pixel23\_2022\_April\_06\_0 to characterize the responsivity of Pixel 23. The dataset comprises repetitive blocks of the diode calibrator in both the ON and OFF states, enabling the subtraction of the background from the pure signal of the calibrator.
By construction, we know that the diode emits $600\pm40$\,K, with uncertainty introduced by the diode switcher (0.1\,dB, which is a Sage SKQ-01 attenuator featuring up to 2.5\,dB insertion loss) and the diode itself (0.2\,dB). The results for each channel are presented in the second row of Table~\ref{tab:pol_angle_total}. The uncertainty is calculated by propagating the uncertainties of the output voltages and the diode temperature in the calculation of responsivity.
%error calculate with uncertainty of 0.3 dB of 600 K

The responsivity calculated from the diode is primarily affected by the uncertainty in the diode's emitted power as seen by the detector, and the fact that this measurement does not reflect the actual observational configuration used on-sky -- for instance, it does not include coupling through the QT2 mirrors, which alters the beam pattern and effective system gain. While the detector’s intrinsic responsivity should, in principle, be independent of the optical setup, the total optical coupling efficiency and beam gain introduced by the optics do influence the system-level responsivity. Additionally, the relatively high temperature of the diode may induce nonlinearities in the detector response, further complicating the calibration.
The calibration diode is mainly used to monitor system responsivity stability rather than to provide an absolute responsivity calibration. However, future measurements will incorporate active thermal control and a lower-output diode to further mitigate possible non-linear effects.

\subsection{Polarization angle calibration}
\label{sec:pol_ang_lab}
%calculated with tfgi_calibration_with_various_angles_polarization_efficiency.py
The reference polarization angle of the channels, $\phi_c$, determines the relation between the measured Stokes $Q$ and $U$ parameters and those associated with the incoming radiation field measured on the Az-El QT2 reference frame.

For this measurement, the zero polarization angle (see \citealp{Tinbergen_1996} for a detailed description) of the entire system for TGI (measured from the MFCI encoder home position), taking into account the telescope (calculating the angle with respect to the vertical) and considering $\epsilon=0$\,deg, was found to be:
\begin{equation}
\rm{ pol\_ang_{TGI} } = 82.5\,\text{deg} + 0.15\,\text{deg} - 90.0\,\text{deg} = -7.35\,\text{deg}
\end{equation}

\begin{figure}[ht]
        \centering
        \includegraphics[width=.45\textwidth]{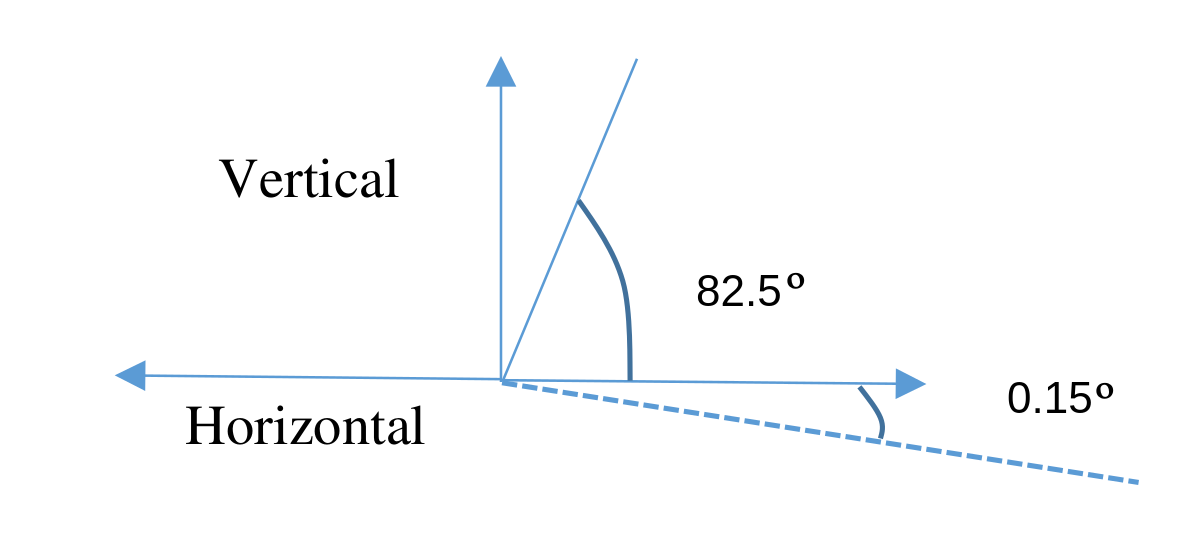}
        \caption{Schematic of the polarization angle measurement of the telescope. The axes represent the ground tilt (Horizontal) and the deviation from Zenith (Vertical).}
        \label{fig:pol_ang_telescope}
\end{figure}

\noindent Figure~\ref{fig:pol_ang_telescope} illustrates the schematic of the zero polarization angle of the telescope, where 0.15\,deg is the horizontal (ground) tilt found when measuring the angle of the fixed plate horizontally of the telescope.  Note that the angles are measured from the horizontal, as a digital inclinometer was utilized, and then referred to the vertical in the equation by subtracting 90\,deg. The double of the resulting angle ($\rm{ pol\_ang_{TGI} }=-7.35$\,deg) is, then, applied to the measurement with diode reference for comparison to the one calculated from astrophysical sources as: 
\begin{equation}
    \phi_{\rm{c}}=\gamma_{\rm{ref}}-2\,\rm{ pol\_ang_{TGI} }
    \label{eq:gamma_ref}
\end{equation}
\noindent where $\gamma_{\rm{ref}}$ is the polarization angle computed in the instrument reference.

Once the telescope's zero polarization angle is calculated, we proceed to characterize the polarization angle using the diode for each channel. The data used for this calculation are the same as described at the beginning of this section.
We calculate the polarization angle as:
\begin{equation}
    \phi_{\text{c}}  = \frac{1}{2} \arctan{\left( \frac{-U}{Q} \right) - 2\,\rm{ pol\_ang_{TGI} } }\text{ ,}
\label{eq:pol_angle_lab}
\end{equation}

\noindent where $U$ and $Q$ are the Stokes parameters computed with the linear combination of the outputs (V) with the equations reported in Appendix~\ref{app:stokes}.
Table~\ref{tab:pol_angle_total} reports the calculated polarization angle for each channel using Eq.~\ref{eq:pol_angle_lab}. The uncertainty is calculated across the different datasets as the standard deviation; the detailed report is in Table~\ref{tab:pol_ang}.

\begin{table}[h]
\caption{Top: Polarization efficiency ($\eta$) for the central pixel (TGI).
Center: Responsivity ($\mathfrak{R}$) with diode reference. Bottom: Polarization angle ($\phi_{\text{c}}$).}
\centering
\small % Makes the text smaller
\begin{tabular}{ccccc}
\hline \hline
  & CH1 & CH2 & CH3 & CH4 \\ \hline
$\eta$ [\%] & $ 96.2 \pm 0.3 $ & $ 91.9 \pm 0.6 $ & $ 96.7 \pm 0.3 $ & $ 94.6 \pm 0.4 $ \\ \hline
$\mathfrak{R}$ [mV/K] & $ 14.2 \pm 1.0 $ & $ 12.9 \pm 1.0 $ & $ 13.4 \pm 1.0 $ & $ 14.5 \pm 1.0 $ \\ \hline
$\phi_{\text{c}}$ [deg] & $131.5 \pm 0.3$ & $-11.7 \pm 0.3$ & $42.9 \pm 0.3$ & $78.8 \pm 0.3$ \\ \hline
\end{tabular}
\label{tab:pol_angle_total}
\end{table}

The difference between polarization angles calculated for consecutive channels is $\sim$48$\pm$9\,deg. This is consistent with the expected value of 45\,deg (note that, as mentioned before, we apply Eq.~\ref{eq:3} for all four channels). The uncertainty ($\pm$9\,deg) is calculated as the standard deviation of the pairwise angular differences, with each channel matched to the nearest other angle in the set. Since the single-angle uncertainty is much smaller, this 9-degree angle seems to correspond to real differences in the angle between channels rather than to a statistical uncertainty.
Appendix~\ref{app:pol_ang_lab} describes the polarization angles for each channel, polarizer angle ($\gamma$), and dataset reported in Table~\ref{tab:pol_ang}. The results in Table~\ref{tab:pol_angle_total} can be compared to those in Table~\ref{tab:gamma_ref_values}, which were obtained using an alternative fitting method explained in the following section.

\subsection{Phase errors calibration}
\label{sec:phase_error_calibration}

To improve the polarization angle reconstruction, we aim to characterize the phase-switch error angles, $\epsilon$, and use them to correct the TFGI equations. For this purpose, we address a 3-parameter problem ($gI$\footnote{$gI$ is considered a single parameter to simplify the system, rather than fitting $g$ and $I$ separately.}, $\epsilon$, $\phi_{\mathrm{c}}$) with six equations corresponding to the six values of $\gamma$, for each channel and phase state. This study considers a distinct $\phi_{\mathrm{c}}$ for each channel. The fitting process unfolds in two phases. Initially, we fit the data to Eq.~\eqref{eq:new_equations} to obtain $\phi_{\rm{c}}$ and $gI$ for each channel without considering any $\epsilon$ correction. Subsequently, with $\phi_{\rm{c}}$ and $gI$ fixed from the previous step, we reintroduce $\epsilon$ into the equations for each channel and phase state.

\paragraph{ $\phi_c$ fitting }
\label{sec:gamma_ref}
The initial fitting step aims to constrain the $\phi_c$ parameter for each channel. We fit Eq.~\eqref{eq:3} under the assumption of $\epsilon = 0$\,deg. Consequently, the resulting $\phi_c$ values encompass the average $\epsilon$ across different phase angles. Subsequently, we perform the fitting for each dataset, with 19, 15, and 15 iterations, respectively, for the three datasets. We exploit the results reported in Table~
\ref{tab:pol_angle_total} (last row) as the initial guess of the fitting for $\phi_c$. The results are reported in Table~\ref{tab:gamma_ref_values}.

\begin{table}[h]
\caption{Fitted polarization angle ($\phi_{\rm{c}}$).}
\centering
\small
\begin{tabular}{ccccc}
\hline \hline
  & CH1 & CH2 & CH3 & CH4 \\ \hline
$\phi_{\text{c}}$ [deg] & $131.3 \pm 0.6$ & $-11.9 \pm 0.6$ & $41.9 \pm 0.6$ & $78.6 \pm 0.6$ \\ \hline
\end{tabular}
\tablefoot{ The results are presented for each channel (CH). The uncertainty is calculated as the standard deviation across the different datasets. }
\label{tab:gamma_ref_values}
\end{table}
%is compatible at the 2-$\sigma$ level with the previous result, as reported in Table~\ref{tab:pol_angle_total}.
The fitting step converges into reference angles for each channel with $\sim$0.6-deg precision, and the final values are compatible at the 2-$\sigma$ level with the initial values reported in Table~\ref{tab:pol_angle_total}.
\paragraph{$\epsilon$ fitting}
We conduct fitting of Eq.~\eqref{eq:new_equations} while maintaining the $\phi_{\text{c}}$ fixed for each channel, as determined in Sect.~\ref{sec:gamma_ref}. This fitting process is carried out individually for each dataset. 
In conclusion, we undertake a statistical analysis by calculating the average values and their corresponding standard deviations across the three datasets. The results are presented in Table~\ref{tab:eps_values}.
%and showcased in Fig.~\ref{fig:epsilon_result}.

\begin{table}[h]
\caption{Phase error corrections ($\epsilon$).}
\centering
\small
\begin{tabular}{ccccc} 
\hline \hline 
\multirow{2}{*}{Channel} & \multicolumn{4}{c}{Error angle correction [deg]} \\
                         & $\delta$=0\,deg & $\delta$=90\,deg & $\delta$=180\,deg & $\delta$=270\,deg \\
\hline 
CH1 & $ 0.9 \pm 1.6 $ & $ 1.5 \pm 0.6 $ & $ 2.2 \pm 1.3 $ & $ 0.6 \pm 1.5 $ \\ 
CH2 & $ -0.2 \pm 1.7 $ & $ 1.0 \pm 1.3 $ & $ 1.9 \pm 0.6 $ & $ 2.4 \pm 1.3 $ \\ 
CH3 & $ 2.3 \pm 1.3 $ & $ 0.7 \pm 1.5 $ & $ 0.8 \pm 1.6 $ & $ 1.4 \pm 0.5 $ \\ 
CH4 & $ 1.6 \pm 0.6 $ & $ 2.1 \pm 1.3 $ & $ 0.3 \pm 1.7 $ & $ 1.1 \pm 1.3 $ \\
\hline
Mean & $1.2\pm0.7$ & $1.3\pm0.6$ & $1.3\pm0.7$ & $1.4\pm0.6$ \\
\hline 
\end{tabular}

\tablefoot{The results are presented for each channel (CH) and phase state ($\delta$). The values represent averages across the datasets used, with the uncertainties calculated as the standard deviation among the different datasets. The final row reports the mean values for each phase state, with uncertainties obtained by propagating the uncertainties in the mean calculation across channels, based on the expectation that the errors should be consistent across channels by design. 
We observe that the quoted uncertainties come from only three datasets tied to this specific instrument setup. Although this limits the statistical strength of the estimates, it reflects the available commissioning data and offers a useful indication of parameter stability; future calibration campaigns will aim to improve this aspect. }
\label{tab:eps_values}
\end{table}
The results are compatible with the null value at the 2$\sigma$ level, where $\sigma$=0.65\,deg corresponds to the mean uncertainty of the mean phase-error values reported in the last row of Table~\ref{tab:eps_values}, which have an average value of 1.3\,deg.
Therefore, we do not need to consider any corrections for scientific purposes at a $\sim$1-degree precision, which is the specification for the TFGI for the polarization angle.
The datasets exploited for the $\epsilon$ calculation are recorded with a separation of $\sim$18\,hours.
%(22-04-06-15-06-58 and 22-04-07-09-48-58), for this study we excluded a dataset which presents a spike
The ambient variables, particularly temperature, may have influenced the response of the phase switches. More datasets at lower timescale differences might be helpful for future development.
The determination of phase-switch errors relies on a non-linear fitting process and can, in theory, depend on the choice of initial parameters or converge to local minima. In practice, incorrect solutions are easily identified through visual inspection of the fitted modulation patterns. Additionally, for data of poor quality, the fit fails to converge to a physically meaningful solution, providing a clear diagnostic for excluding unreliable measurements. These checks help prevent biases in estimating the phase-switch errors.
Appendix~\ref{app:fitting_model} presents the results per dataset.

\section{On-sky calibration}
\label{sec:on_sky_calibration}

After calibrating the instrument using the diode reference on the ground (as described in the previous section), we characterize the instrument with astrophysical sources. We perform various observing tests during the instrument's commissioning phase, employing observations of Tau A (see Sect.~\ref{sec:cal_TauA}) and the Moon (see Sect.~\ref{sec:cal_moon}) to characterize TGI Pixel 23. The following sections detail the calibration methods used and present the respective results, which will be compared across all methods and calibration sources studied in this work in Sect.~\ref{sec:comparison}.

\subsection{Tau A}
\label{sec:cal_TauA}

\subsubsection{Polarization efficiency with Tau A}
\label{sec:pol_eff_tauA}
We use Tau A observations to estimate the polarization efficiency similarly to that described in Section~\ref{sec:pol_eff_lab}. Tau A has been extensively used for this purpose in the past \citep[e.g.,][]{mfiwidesurvey} thanks to being very bright at radio-to-microwave frequencies ($>300$\,Jy at frequencies $<10$\,GHz). It presents a high polarization fraction, being $\sim$7\% at frequencies $>10$\,GHz \citep[e.g.,][]{ritacco_crab}.

\begin{figure*}
    \centering
    \includegraphics[width=.9\linewidth]{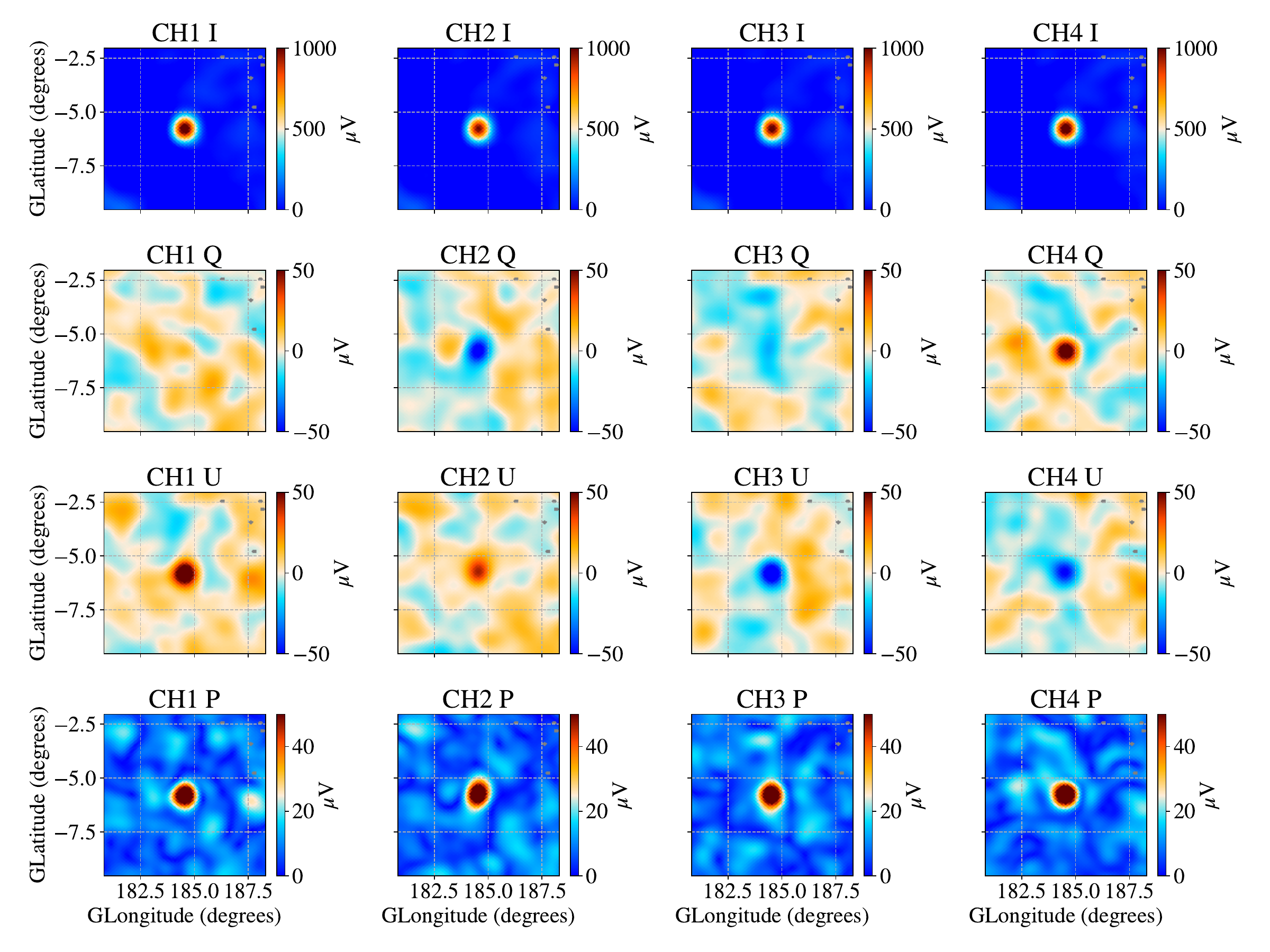}
    \caption{Maps of the single Tau A observation used to estimate polarization efficiency, smoothed at 1\,deg. We discuss in Sect.~\ref{sec:pol_ang_crab} how these uncalibrated maps are utilized to calculate the polarization angle. Rows, from top to bottom: Stokes parameters $I$, $Q$, $U$, and polarized intensity $P$. Columns, from left to right: channels 1, 2, 3, and 4.}
    \label{fig:crab_single_obs_maps}
\end{figure*}

For this analysis, we selected the Tau A observation with the best noise properties while being close in time to the Moon observation that will be discussed in Sect.~\ref{sec:cal_moon}. This observation consists of a single raster scan, and its properties are summarized in Table~\ref{tab:TauA_211130}.
From this observation, we produce maps of the three Stokes parameters for each channel independently, which %. These maps are later smoothed to a 1\,deg FWHM Gaussian beam, and 
are shown in Fig.~\ref{fig:crab_single_obs_maps}. The detection of a polarized signal from Tau A is immediately clear. To estimate the source flux density, we use a beam-fitting (BF) photometry that involves fitting the $I$, $Q$, and $U$ maps of each channel to a 2D Gaussian with fixed position (using the nominal coordinates of Tau A) and width fixed to the FWHM of the beam calculated in \cite{tfgi_commissioning}. The amplitudes of the Gaussians obtained from these fits define $I_{\rm BF}$, $Q_{\rm BF}$ and $U_{\rm BF}$ estimates, with their uncertainties being $\sigma(I_{\rm BF})$, $\sigma(Q_{\rm BF})$ and $\sigma(U_{\rm BF})$.
At the TFGI angular resolution, Tau A is only marginally resolved and is therefore well described by a Gaussian profile; any residual source-extension effects are absorbed into the BF uncertainty.
As a consistency check, we also performed aperture photometry on the Tau A observations and obtained results consistent with the BF analysis.

Polarized intensity, $P$, maps in Fig.~\ref{fig:crab_single_obs_maps} are calculated as $\sqrt{Q^2 + U^2}$ for visualization purposes. However, the estimates for $P$ used to compute the polarization efficiency are not directly derived from the $P$ maps; rather, they come from the $Q$ and $U$ BF estimates, $Q_{\rm BF}$ and $U_{\rm BF}$. We divide this $P$ by the fitted $I_{\rm BG}$ and compare the result with the model. This approach minimizes the noise bias that affects polarized intensity measurements in regions with low signal-to-noise ratios.

\begin{table}[]
    \caption{Raster scan observation properties and characterization of Pixel 23 from Tau A. The upper part lists the scan parameters, while the lower part reports polarization efficiency ($\eta$) and responsivity ($\mathfrak{R}$) for each channel (CH).}
    \centering
    \small
        \begin{tabular}{cc}
            \hline \hline
            Parameter & Value \\
            \hline
            Observation date & 2021-11-30, 02:30 UTC \\
            Raster size (RA, Dec) [deg] & $16.1 \times 17.6$ \\
            Exposure time [hours] & 1.09 \\
            %AP radii [deg] & 1, $\sqrt{2}$, 2 \\
            Flux density, $S$ @ 31\,GHz [Jy] & 323.23 \\
            Temperature, $T$ @ 31\,GHz [mK] & 230 \\
            \hline
            \multicolumn{1}{c}{Channel} & \multicolumn{1}{c}{$\eta$ [\%] \quad\quad $\mathfrak{R}$ [mV/K]} \\
            \hline
            \multicolumn{1}{c}{CH1} & $91 \pm 4$ \quad\quad $37.3 \pm 0.6$ \\
            \multicolumn{1}{c}{CH2} & $98 \pm 4$ \quad\quad $33.5 \pm 0.5$ \\
            \multicolumn{1}{c}{CH3} & $97 \pm 6$ \quad\quad $34.5 \pm 0.5$ \\
            \multicolumn{1}{c}{CH4} & $93 \pm 6$ \quad\quad $38.0 \pm 0.6$ \\
            \hline
        \end{tabular}
    \tablefoot{Single Tau A observation conducted on 30 November 2021.
    We chose this scan as it showed one of the lowest noise estimates of our sample, and it was close in time to the Moon observations from Section~\ref{sec:cal_moon}.
    Additional scans are included in the monthly coadded analysis presented in Table~\ref{tab:all_tauA_monthly_gains} and will be further discussed in \citealp{tfgi_commissioning}}.
    \label{tab:TauA_211130}
\end{table}

We use the ratio of the measured polarization fraction to the expected value from previous measurements to estimate the polarization efficiency of our detectors. We assumed the expected value would be constant and equal to $7\pm0.1$\,\% at TFGI frequencies.
We obtained this value from the weighted average of the results for the five frequency bands given in \cite{weiland2011}. 
Although the FWHM values in \cite{weiland2011} bands are larger than the one from TFGI, the signal from Tau A is fully contained in one beam in both cases. Because of that, we do not expect a change in the polarization fraction, as shown by previous works (see e.g. Fig.~39 from \citealt{mfiwidesurvey}).
We did not propagate this 0.1\% uncertainty to the final map measurements, as the errors in $IQU$ measurements dominate. We find an average polarization efficiency across the channels of $\eta=95\pm3$\% (the uncertainty is calculated as the standard deviation across the channels), with a single-channel efficiency between $\eta\sim$91--98\,\% (as seen in Table\,\ref{tab:TauA_211130}).

\subsubsection{Responsivity calibration with Tau A}
\label{sec:responsivity_tauA}

The estimates obtained in the previous section with BF are compared with the expected values derived from the Tau A model presented in \cite{mfiwidesurvey} in order to estimate its responsivity:
\begin{equation}
    S_\nu = 358.3\,{\rm Jy}\left(\frac{\nu}{22.8\,{\rm GHz}}\right)^{-0.297} \text{ .}
    \label{eq:crab_flux_model}
\end{equation}

\noindent We apply the $-0.21\%\,{\rm year}^{-1}$ secular decrease value from 
\cite{weiland2011}, using $t_0=2016.3$\,year as the reference time for the model and $t=2021.9$\,year for the observation\footnote{This translates into a decrease of the flux density by a factor $\exp\left[{\frac{-0.21}{100}\,(2021.9-2016.3)}\right]=0.9883$.}. The expected flux is therefore $S_{31\,{\rm GHz}}=323.23\,$Jy. We convert this into temperature  as\footnote{We consider only the nominal frequencies of TFGI, 31 and 41\,GHz. Color corrections have been simulated assuming a top-hat band and are found to be negligible ($<$1\%), so they are not applied.}:
\begin{equation}
    % T_\nu = S_\nu \left(\frac{2k_{\rm B}\nu^2}{c^2}\frac{x^2e^x}{(e^x-1)^2}\Omega_b\right)^{-1} \text{ ,}
    T_\nu = S_\nu \left(\frac{2k_{\rm B}\nu^2}{c^2}\Omega_b\right)^{-1} \text{ ,}
\end{equation}

\noindent where $\Omega_b$ is the solid angle subtended by the model beam, 
%$\Omega_b= 4.23\times10^{-5}~\mathrm{sr}$ %this is with 21 arcmin FWHM
$\Omega_b= 4.77\times10^{-5}~\mathrm{sr}$. %with 22.3 arcmin FWHM
The expected temperature at 31\,GHz is then 230\,mK.
We obtain the responsivity by dividing the amplitude calculated through BF photometry on the intensity map, $I_{\rm BF}$, by the calculated temperature.
We report the responsivities derived from Tau A in Table~\ref{tab:TauA_211130} (for the single raster scan described in the previous section).
%, and convert it to Rayleigh-Jeans temperature \footnote{By multiplying by a factor $x^2\times e^x/(e^x-1)^2\approx0.976$.},
%nu = 31e9
%x = h*nu/k/2.7255
%C = x**2.*np.exp(x)/(np.exp(x)-1.)**2.
We also calculated the responsivity values for the coadded monthly maps (from \citealp{tfgi_commissioning}) from November 2021, December 2021, and January 2022. 
We report the responsivities derived from Tau A in Tables~\ref{tab:TauA_211130} (for the single raster scan described in the previous section) and \ref{tab:all_tauA_monthly_gains} (for the monthly coadded maps). We also compute the responsivity from Moon observations in the following Section; we present this sky-sky check of the responsivity in Sect.~\ref{sec:comparison_responsivity}.

\begin{table}[]
    \caption{Monthly estimates of the responsivity ($\mathfrak{R}$) for each channel (CH) computed from Tau A. We present how these maps were produced and the computation of their uncertainties in \citealt{tfgi_commissioning}.}
    \centering
    \small
    \begin{tabular}{rcccc}
\hline \hline
\multirow{2}{*}{Month}  &    \multicolumn{4}{c}{Responsivity [mV/K]} \\
  & CH1     &    CH2     &    CH3     &    CH4     \\
\hline
  Nov 21  & $36.5\pm0.8$ & $33.2\pm0.6$ & $34.4\pm0.9$ & $37.5\pm1.0$ \\
  Dec 21  & $35.9\pm0.2$ & $32.1\pm0.5$ & $33.8\pm0.3$ & $36.7\pm0.8$ \\
  Jan 22  & $26.3\pm0.6$ & $23.9\pm0.6$ & $25.1\pm0.6$ & $27.4\pm0.7$ \\
\hline
\end{tabular}

% 21 November 2021
% 28 December 2021
% 23 January  2022
    \tablefoot{Observations conducted from November 2021 to January 2022.}
    \label{tab:all_tauA_monthly_gains}
\end{table}

\subsubsection{Polarization angle}
\label{sec:pol_ang_crab}

We compute the reference polarization angle from the comparison between the one calculated from our raw on-sky data, $\gamma'$, and the expected one, $\gamma_{\rm Tau A}$:
\begin{equation}
    \phi_c = \gamma_{\rm Tau A} - \gamma' = \gamma_0 + {\rm RM}\lambda^2 - \frac{1}{2}\arctan
    \left(\frac{-U}{Q}\right)
\label{eq:pol_angle_sky}
\end{equation}
\noindent where $\gamma_0=-88.31\pm0.25\,\deg$ and ${\rm RM} = -1\,406\pm12\,\deg\,{\rm m}^{-2}$ is the rotation measure (\citealp{mfiwidesurvey}). 
Additionally, the minus sign for $\gamma'$ represents the change in the x-axis direction when transitioning from the telescope to the sky coordinate system.

We compute $\phi_c$ from $Q_{\rm BF}$ and $U_{\rm BF}$ values from Sect.~\ref{sec:pol_eff_tauA} using
Eq.~\eqref{eq:pol_angle_sky}. We summarize the results in Table~\ref{tab:polar_angle_crab}. Finally, we estimate the uncertainties of $\phi_{\rm{c}}$ by drawing 10\,000 random realizations for $Q$ and $U$ distributions. We define these distributions as Gaussians with centers equal to $Q_{\rm BF}$ and $U_{\rm BF}$ and standard deviations equal to $\sigma(Q_{\rm BF})$ and $\sigma(U_{\rm BF})$.
% \textbf{from Gaussian distributions with standard deviations given by the BF uncertainties.} 
We evaluated $\phi_{\rm c}$ for each pair $Q_{\rm BF}-U_{\rm BF}$ and then built its distribution. We defined the uncertainty on $\phi_{\rm c}$ as the width of such a distribution.
% The width of the resulting $\phi_{\rm c}$ distribution was used to estimate the uncertainty of this parameter.

\begin{table}[]
    \caption{Polarization angles ($\phi_{\rm{c}}$) for each channel (CH) derived from the Tau A observation shown in Fig.~\ref{fig:crab_single_obs_maps}.}
    \centering
    \small
    \begin{tabular}{ccccc}
\hline \hline
 & CH1 & CH2 & CH3 & CH4 \\ \hline
% $\phi_c$ [deg] & $140.2 \pm 9.2$ & $-14.4 \pm 7.4$ & $34.2 \pm 9.6$ & $65.2 \pm 11.9$ \\ \hline
$\phi_c$ [deg] & $132.0 \pm 1.4$ & $-15.1 \pm 1.3$ & $48.7 \pm 1.4$ & $73.9 \pm 1.5$ \\ \hline
\end{tabular}
    \label{tab:polar_angle_crab}
\end{table}

\subsection{The Moon}
\label{sec:cal_moon}

% 21 November 2021
% 28 December 2021
% 23 January  2022

We use three Moon observations to characterize TGI Pixel 23 in the sky. 
These observations were performed on 21 November 2021, 28 December 2021, and 23 January 2022.
Each observation involves repeated azimuth slews (azimuth scans at fixed elevation) lasting $\sim$1\,hour, covering an azimuth range of 100\,deg at a speed of 4.5\,deg/s.
The azimuth scan spans 100\,deg to ensure full beam crossing for all pixels, thereby minimizing edge effects and ensuring uniform beam sampling.

\subsubsection{Moon emission model}

Moon's polarized microwave emission is produced mainly by its own thermal emission. Its brightness temperature can be modeled, following \cite{Hafez_2014}, as:
\begin{equation}
    T_\mathrm{Moon} = T_0(\nu) + T_1(\nu) \cos{\left(\phi(t) - \xi(\nu)\right)},
    \label{eq:Moon:Thermal}
\end{equation}
\noindent where $T_0$ represents the mean temperature and $T_1$ is the amplitude of the temperature variation associated with the Moon's phase, $\phi(t)$. Finally, $\xi(\nu)$ is the phase relative to the time of the full Moon. The values at 31\,GHz for these parameters are interpolated from the values presented in \cite{Hafez_2014}, obtaining: $T_0(31\,\rm{GHz}) = 213.3 \pm 0.9$\,K, $T_1(31\,\rm{GHz}) = 35 \pm 2$\,K $\xi(31\,\rm{GHz}) = 28\pm 4$\,deg.
For our three observing dates (21 November 2021, 28 December 2021, and 23 January 2022), the corresponding lunar phases are $\phi$=19.4\,deg, 104.4\,deg, and 60.2\,deg, respectively.
In this work, we do not include spatial dependence on the thermal signal due to the low angular resolution of the telescope, and it is not expected to significantly affect either the absolute signal (at most $\sim$8\% discrepancy with \citealp{2012Icar..219..194Z} model in these dates) or the polarization morphology. Additionally, our observations were made relatively near the time of the full Moon, which also decreases the contribution of spatial anisotropies.

\begin{figure}[ht]
    \centering
    \includegraphics[width = 0.4\textwidth]{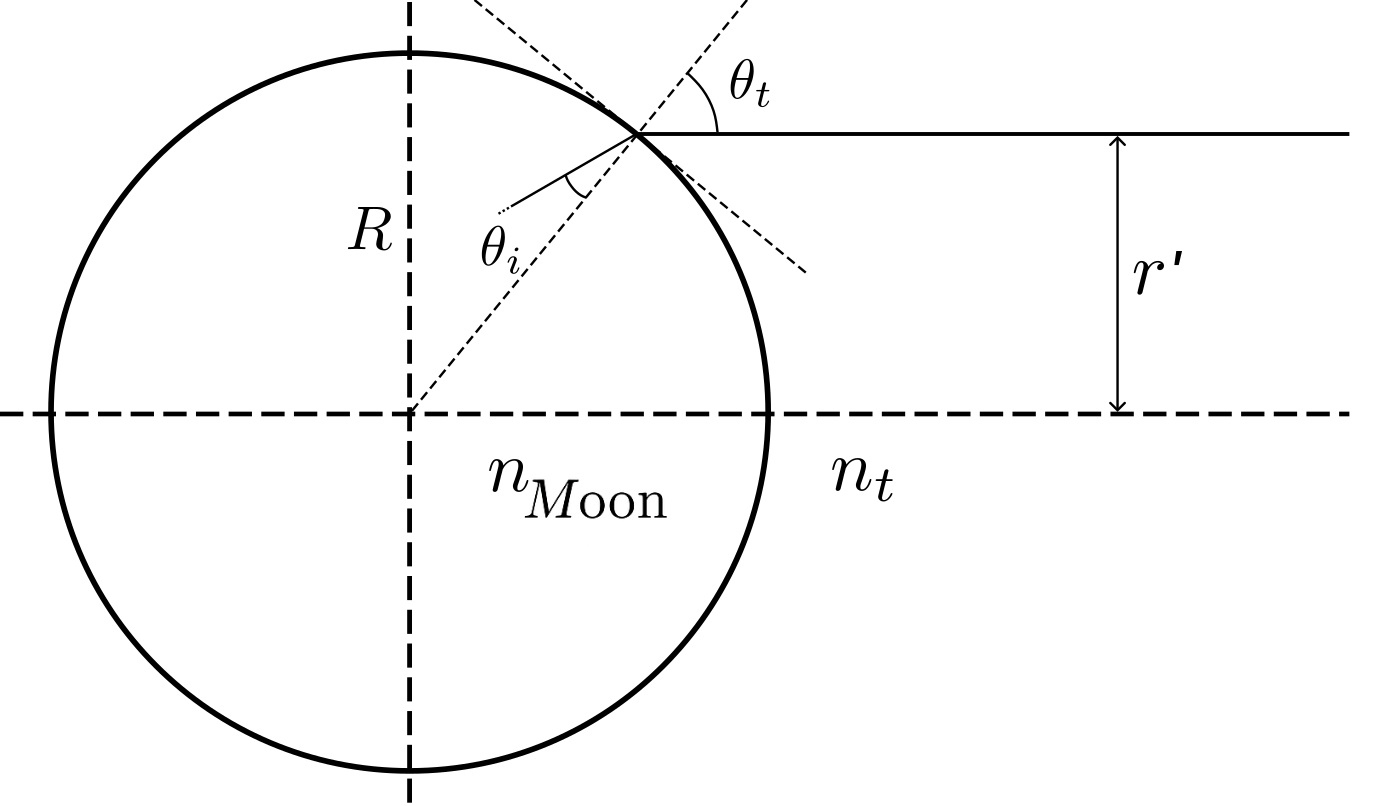}
    \caption{Schematic describing the Moon's emission model and the parameters involved: the refraction index of the Moon ($n_{\rm{Moon}}$), the refraction index of free space ($n_t$), the incident ($\theta_i$) and transmitted ($\theta_t$) angles, the radial coordinate ($r'$), and the radius of the Moon ($R$).}
    \label{fig:Moon:MoonEmission}
\end{figure}

Additionally, the modeling of polarized emission is feasible, as it is only weakly polarized because of the refraction of thermal radiation on the Moon's surface.
Following Fresnel equations (see, e.g., \citealp{1959pot1.book.....B}), the transmittances of the perpendicular $t_\perp$ and the parallel $t_\parallel$ components are defined:
\begin{equation}
    \begin{array}{lcl}
         t_{\perp} & = & \dfrac{2 n_{\rm{Moon}}\cos{\theta_i}}{n_{\rm{Moon}}\cos{\theta_i} + n_t\cos{\theta_t}},\\
         \\
         t_{\parallel} & = & \dfrac{2 n_{\rm{Moon}}\cos{\theta_i}}{n_t\cos{\theta_i} + n_{\rm{Moon}}\cos{\theta_t}},\\
    \end{array}
    \label{eq:Moon:Fresnel}
\end{equation}

\noindent where $n_{\rm{Moon}}$ is the refraction index of the Moon, $n_t$ is the refraction index of the free space ($n_t \approx 1$), and $\theta_i$ and $\theta_t$ are the incident and transmitted angles, respectively. To better understand these angles, we refer to Fig.~\ref{fig:Moon:MoonEmission}, from which we can derive the relation $\theta_t = \arcsin{(r'/R)}$, where $r'$ is the radial coordinate and $R$ is the radius of the Moon.

\noindent By exploiting the transmittance values described in Eqs.~\eqref{eq:Moon:Fresnel}, we calculate the intensity ($I_{\rm{Moon}}$) and the polarization ($P_{\rm{Moon}}$) of the Moon's surface as a function of its temperature ($T_{\mathrm{Moon}}$):
\begin{equation}
     \begin{array}{lcl}
         I_{\rm{Moon}} & = & \dfrac{T_\mathrm{Moon}}{2} \left( t_\parallel^2  + t_\perp^2 \right) \dfrac{\partial\theta_i}{\partial\theta_t},\\
         \\
         P_{\rm{Moon}} & = & \dfrac{T_\mathrm{Moon}}{2} \left(t_\parallel^2 - t_\perp^2 \right) \dfrac{\partial\theta_i}{\partial\theta_t},\\
    \end{array}
    \label{eq:Moon:T}
\end{equation}
\noindent where the term $\partial\theta_i/\partial\theta_t$ describes the change in the solid angle of the incident beam after the refraction process.

By solving these equations, we obtain the radial intensity and polarization patterns for the Moon's emission. Notice that the polarization direction is parallel to the radial direction, and then it has rotational symmetry. Therefore, if we consider that in a reference frame centered on the Moon the polarization signal is entirely contained in the Stokes $Q$ parameter, we can estimate the Stokes parameters $Q_{\rm p}$,  $U_{\rm p}$, of a given pixel $p$ by applying the following rotation:
\begin{equation}
    \begin{pmatrix} Q_p\\ U_p \end{pmatrix} =
    \begin{pmatrix}
        \cos{2\varphi_p} & -\sin{2\varphi_p}\\
        \sin{2\varphi_p} & \cos{2\varphi_p}\\
    \end{pmatrix} \begin{pmatrix} P_{\rm{Moon},\,p} \\ 0 \end{pmatrix},
\end{equation}
\noindent where $\varphi_{p}$ is the polar angle that gives the orientation of the pixel $p$ with respect to the map axis (North-East direction).

Thus, using this model, we can obtain the Stokes parameters in the North-East Moon-centered coordinate frame and the total polarization intensity $P_{\rm{tot}} = \sqrt{Q^2+U^2}$ as a function of $T_\mathrm{Moon}$ and $n_{\rm{Moon}}$. 

\begin{figure}
    \centering    
    \includegraphics[width=0.4\textwidth]{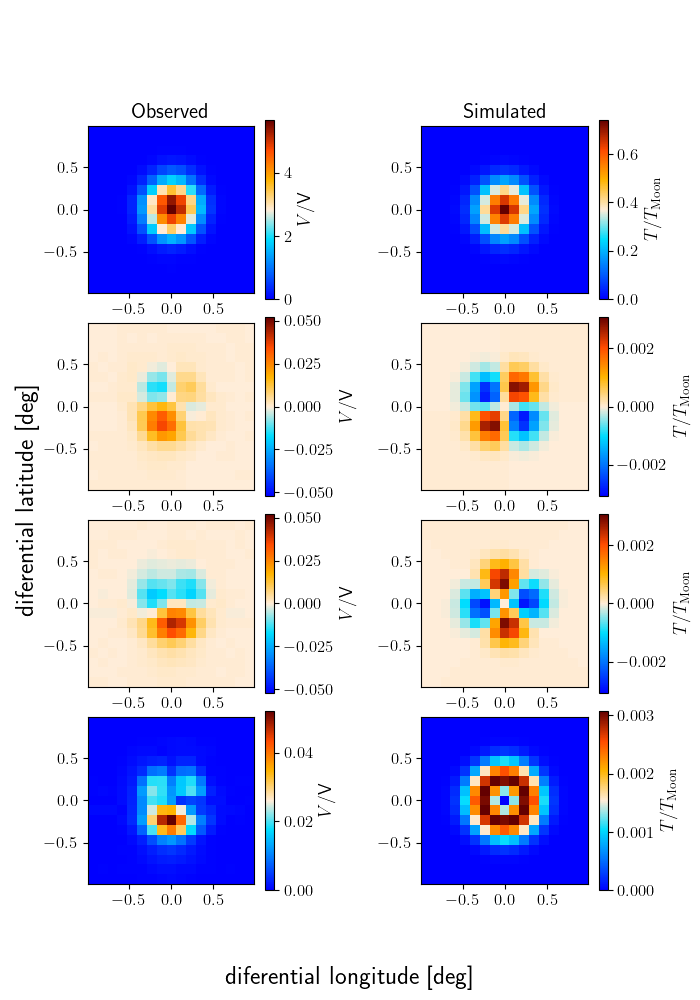}
    \caption{Moon maps of the three Stokes parameters $I$, $Q$, and $U$ and of the polarized intensity $P$ for an observation taken on December 2021 (left), and for the simulated model inferred from the observation parameters (right).}
    \label{fig:Moon:MoonObsvSim}
\end{figure}

Finally, to compare the model with the measured data and calibrate the various parameters of our instrument, we also need to consider the beam shape of our instrument. First, we convolve the modeled maps with a Gaussian beam with the TGI's FWHM (21\,arcmin). After that, as it will be described in Sect.~\ref{sec:pol_ang_moon}, we find the relative angle that gives the best fit between the model and the data. 
Figure~\ref{fig:Moon:MoonObsvSim} displays both the observed (left) and simulated (right) maps using a fixed refraction index $n_{\rm Moon}=1.22$ (from \citealp{1967SvA....11..329L} at 37\,GHz), after applying all the corrections (for a detailed description of this model, see e.g., \citealp{bischoff_observing_2010, perley_integrated_2013, xu_two-year_2020, dahal_microwave_2023}).

As in \cite{2020ApJ...891..134X}, we do not observe the emergence of the quadrupolar polarization pattern in a single observation of the Moon for the central detector. In this work, we assume a uniform temperature, $T_{\rm{Moon}}$, across the lunar disk, thereby simplifying the latitude-dependent model proposed by \citet{2012Icar..219..194Z}. Even after including the latitude-dependent model and stacking the three available maps taken at different epochs
%\footnote{27 December 2021,ade2019simons 28 December 2021, and 13 February 2022} 
and incorporating the latitude-dependent temperature model, the quadrupolar pattern is not recovered in Pixel 23 (central pixel). The behavior is repeatable and stable over the timescale of our observations ($\sim$2 months). In contrast, the quadrupolar structure is clearly visible in single observations from other TFGI detectors.
We performed preliminary simulations and tests to investigate potential explanations, including intensity-to-polarization leakage, non-linear response, saturation effects, phase state angle misestimation, and mapping strategy artifacts. None of these effects explains the observed behavior. Therefore, the origin of this discrepancy for the central detector remains unclear. 
Even when the quadrupolar pattern is only partially visible (i.e., one lobe is missing), the relative orientation between the polarization patterns observed in Stokes $Q$ and $U$ is preserved. Since the polarization angle determination relies on these relative rotations rather than on the full quadrupolar morphology, the measurements still provide a robust estimate of the polarization angle.
A dedicated future study will address this issue in detail, including a comprehensive analysis of Moon observations across all TFGI detectors, once the instrument is installed again on the focal plane of QT2.
The TFGI instrument was temporarily removed after the initial commissioning phase. 
A new configuration comprising 19 receivers (10 at 31\,GHz and 9 at 41\,GHz) is being prepared, and observations are expected to resume in spring 2026.

\subsubsection{Polarization efficiency with the Moon}
\label{sec:pol_eff_moon}

We recover the polarization efficiency by comparing the polarization fractions obtained from the measurements with those from the modeled signal.
The modeled polarization fraction depends tightly on the refraction index of the Moon, $n_{\rm{Moon}}$, which has not been directly measured with the required precision to date at 31\,GHz.

In a first attempt, we derived an estimate of the polarization efficiency as $\eta=P_{\rm{obs}}/P_{\rm{mod}}$ %pol = Pi_obs/Pi_mod
where $P_{i} = \sqrt{Q_i^2+U_i^2}/I_i$
and $Q_i$, $U_i$, and $I_i$ are the flux densities calculated, through an aperture photometry (AP) technique with a radius of 48\,arcmin, on the corresponding maps resulting from the observing data ($i$=obs) or from the model ($i$=mod).
We assumed a value of the refraction index $n_{\rm{Moon}}=1.22$ (from \citealp{1967SvA....11..329L}) and obtained the results reported in Table~\ref{tab:Moon:Peff} for each channel and for each observation.
The mean polarization efficiency across channels and observations is $\eta=89\pm2$\%, with the uncertainty calculated as the standard deviation across the mean values for each channel.

\begin{table}
    \caption{Polarization efficiencies ($\eta$) for each channel (CH) derived from the Moon.}
    \centering
    \small % or \footnotesize
    \begin{tabular}{rcccc}
        \hline \hline
        \multirow{2}{*}{Month}  &    \multicolumn{4}{c}{Polarization efficiency [\%]} \\
        & CH1     &    CH2     &    CH3     &    CH4     \\
        \hline
        Nov 21 & 86 $\pm$ 27 & 85 $\pm$ 26 & 88 $\pm$ 27 & 86 $\pm$ 27 \\
        Dec 21 & 93 $\pm$ 29 & 91 $\pm$ 28 & 93 $\pm$ 29 & 88 $\pm$ 27 \\
        Jan 22 & 88 $\pm$ 28 & 88 $\pm$ 27 & 91 $\pm$ 28 & 85 $\pm$ 26 \\
        \hline
        Mean & 89 $\pm$ 16 & 88 $\pm$ 16 & 91 $\pm$ 16 & 86 $\pm$ 15 \\
        \hline
    \end{tabular}
    \tablefoot{The last row presents the average value for each channel across the observations, with the uncertainty calculated by propagating the errors in the mean calculation.}
    \label{tab:Moon:Peff}
\end{table}

It is important to note at this point that the uncertainty on the polarization efficiency recovered value is mainly dominated by the uncertainty on the refraction index, which has been set as 0.05.
Given the difficulty of assessing the exact value of $n_{\rm Moon}$, and of its uncertainty, due to the lack of measurements in this frequency range, we have decided to invert the problem and derive a value of $n_{\rm Moon}$ from our observations by fixing the polarization efficiency to the values derived from Tau A shown in Table~\ref{tab:TauA_211130}. This approach, which is presented in Appendix~\ref{sec:moon_refraction_index}, yields an effective refraction index of $n_{\rm{Moon}} = 1.209 \pm 0.007\, \text{(stat)}\,\pm0.009\,\text{(sys)}$, compatible at $\sim$1$\sigma$ with the value constrained by \cite{1967SvA....11..329L}.

\subsubsection{Responsivity calibration with the Moon}
\label{sec:responsivity_moon}

We calculate the Moon's brightness temperature from Eq.~\eqref{eq:Moon:Thermal}.
We can, thus, model the thermal emission of the Moon at 31\,GHz on different dates (Moon phases) and simulate, similar to the previous sections, emission maps that we directly compare with the observations. We perform an AP study on both simulated and measured maps in intensity (e.g., as in the first row of Fig.~\ref{fig:Moon:MoonObsvSim}) and compare them to determine the detector's responsivity. The calculated responsivities for Pixel 23 are reported in Table~\ref{tab:Moon:GainCal}.
We compare the Moon and Tau A results in Sect.~\ref{sec:comparison_responsivity}.
\begin{table}
    \caption{Responsivity ($\mathfrak{R}$) for each channel (CH) derived from three Moon observations.}
    \centering
    \small
    \begin{tabular}{rcccc}
\hline \hline
\multirow{2}{*}{Month}  &    \multicolumn{4}{c}{Responsivity [mV/K]} \\
  & CH1     &    CH2     &    CH3     &    CH4     \\
\hline
Nov 21 & 33.6 $\pm$ 0.3 & 30.4 $\pm$ 0.3 & 31.6 $\pm$ 0.4 & 34.3 $\pm$ 0.4 \\
Dec 21 & 38.7 $\pm$ 0.4 & 34.7 $\pm$ 0.4 & 36.6 $\pm$ 0.4 & 39.4 $\pm$ 0.4 \\
Jan 22 & 28.4 $\pm$ 0.3 & 25.7 $\pm$ 0.3 & 26.6 $\pm$ 0.3 & 29.0 $\pm$ 0.3 \\

\hline
\end{tabular}

    \tablefoot{The errors presented on this table are obtained by propagating the error arising from the Moon temperature model presented on \cite{Hafez_2014} (eq. \ref{eq:Moon:Thermal}).}
    \label{tab:Moon:GainCal}
\end{table}

\subsubsection{Polarization angle}
\label{sec:pol_ang_moon}

The polarized emission of the Moon exhibits a characteristic quadrupolar pattern, as illustrated in the model shown on the right of Fig.~\ref{fig:Moon:MoonObsvSim}. When observed with sufficient angular resolution to resolve this structure, the polarization angle of the detectors ($\phi_{\rm{c}}$) can be determined by fitting the real data to the model. Specifically, the Stokes parameters measured by the instrument ($Q_{\rm obs}$ and $U_{\rm obs}$) are related to the modeled Stokes parameters ($Q$ and $U$) through a rotation by an angle of $2\phi_{\rm{c}}$, as:
\begin{equation}
    \begin{pmatrix}
        Q\\
        U\\
    \end{pmatrix} = \begin{pmatrix}
        \cos{2\phi_{\rm{c}}} & -\sin{2\phi_{\rm{c}}}\\
        \sin{2\phi_{\rm{c}}} & \cos{2\phi_{\rm{c}}}\\
    \end{pmatrix} \begin{pmatrix}
        Q_{\rm obs}\\
        U_{\rm obs}\\
    \end{pmatrix}.
    \label{eq:Moon:gammaref_1}
\end{equation}

\noindent The fitting is executed by minimizing the residuals (i.e., the $\chi^2$ of the fit) between the data and the model evaluated for a given angle $\phi_{\rm c}$:
\begin{align}
    \chi^2(\phi_{\rm{c}}) =& \frac{1}{\sigma_{\rm{DET}}^2}\sum_p \bigg[\left( \dfrac{Q_{\rm obs, \textit{p}}}{\max(I_{\rm obs})} - \dfrac{Q_{p}(2\phi_{\rm{c}})}{\max(I)} \right)^2  +      \\
    &+ \left( \dfrac{U_{\rm obs, \textit{p}}}{\max(I_{\rm obs})} - \dfrac{U_{p}(2\phi_{\rm{c}})}{\max(I)} \right)^2 \bigg]. \nonumber
        \label{eq:Moon:minphi}
\end{align}

\noindent where the sum runs over all pixels $p$ inside an aperture of radius 1\,deg, $\sigma_{\rm DET}$ is is the detector's noise, $\max(I)$ and $\max(I_{\rm obs})$ are the maximum intensity values measured on the respective maps and $Q_{p}(2\phi_{\rm{c}})$ and $U_{p}(2\phi_{\rm{c}})$ are the $2\phi_{\rm{c}}$ rotations of the modeled maps:
\begin{equation}
    \begin{pmatrix}
        Q(2\phi_{\rm{c}})\\
        U(2\phi_{\rm{c}})\\
    \end{pmatrix} = \begin{pmatrix}
        \cos{2\phi_{\rm{c}}} & -\sin{2\phi_{\rm{c}}}\\
        \sin{2\phi_{\rm{c}}} & \cos{2\phi_{\rm{c}}}\\
    \end{pmatrix} \begin{pmatrix}
        Q\\
        U\\
    \end{pmatrix}.
    \label{eq:Moon:gammaref_2}
\end{equation}

Considering the spatial properties of this fit, it is crucial to create simulated maps that exhibit characteristics similar to the observations, as shown in Fig.~\ref{fig:Moon:MoonObsvSim}, where this fitting method has already been applied.
The best-fit values for the polarization angle obtained using this method are shown in Table~\ref{tab:Moon:PolAngle}, where the uncertainties correspond to those extracted from the fitting procedure. We observe differences of up to $\sim$3\,deg between the three observations, which are distributed over a period of three months.

Finally, we note that the resulting fitting error is small, with a typical uncertainty of $0.04$\,deg. However, this estimate does not include systematic effects, which are expected to dominate the total error budget.
Model-related uncertainties, including those linked to the latitude-dependent temperature model, are not considered in this work. Preliminary simulations indicate that these effects do not significantly influence the results; however, a complete evaluation of such model uncertainties is beyond the scope of this work and will be addressed in future dedicated studies.
Overall, the methodology appears promising; with further development, it could become a precise tool for characterizing the polarization angle in ground-based experiments.

\begin{table*}
    \caption{Polarization angles ($\phi_{\rm{c}}$) for each channel (CH) derived from three Moons observations.}
    \centering
    \small
    \begin{tabular}{rcccc}
\hline \hline
\multirow{2}{*}{Month}  &    \multicolumn{4}{c}{Polarization angle [deg]}  \\
  & CH1     &    CH2     &    CH3     &    CH4     \\
\hline

Nov 21 & 124.86 $\pm$ 0.03 & $-18.66$ $\pm$ 0.04 & 36.49 $\pm$ 0.03 & 72.74 $\pm$ 0.04 \\
Dec 21 & 128.16 $\pm$ 0.05 & $-15.37$ $\pm$ 0.04 & 38.91 $\pm$ 0.04 & 75.33 $\pm$ 0.04 \\
Jan 22 & 130.06 $\pm$ 0.10 & $-12.49$ $\pm$ 0.05 & 41.56 $\pm$ 0.04 & 76.97 $\pm$ 0.04 \\

%Nov 21 & $ 124.8 \pm 0.2 $ & $ -18.6 \pm 0.1 $ & $ 36.4 \pm 0.6 $ & $ 72.6 \pm 1.0 $ \\
%Dec 21 & $ 128.1 \pm 0.2 $ & $ -15.3 \pm 0.1 $ & $ 39.0 \pm 0.5 $ & $ 75.3 \pm 0.9 $ \\
%Jan 22 & $ 130.1 \pm 0.2 $ & $ -12.5 \pm 0.1 $ & $ 41.4 \pm 0.6 $ & $ 76.9 \pm 0.9 $ \\
%Sep 22 & $ 75.58 \pm 8.99 $ & $ 76.6 \pm 9.08 $ & $ 83.01 \pm 8.38 $ & $ 79.67 \pm 9.21 $ \\
\hline
Mean & $127.2 \pm 2.2$ & $-15.5 \pm 2.5$ & $38.9 \pm 2.1$ & $75.1 \pm 1.7$ \\

\hline
\end{tabular}

    \tablefoot{The uncertainty for each observation comes from the fitting uncertainty. The last row presents the average value for each channel across the observations, and the uncertainties are calculated as the standard deviation of the measurements.}
    \label{tab:Moon:PolAngle}
\end{table*}

\section{Comparison of the different methods' results}
\label{sec:comparison}

\subsection{Polarization efficiency}

Table~\ref{tab:pol_angle} summarizes the polarization efficiencies calculated with the MFCI, Tau A, and the Moon setups for each channel. 
%Section~\ref{sec:pol_eff_lab} details the results from ground, refer to Sect.~\ref{sec:pol_eff_tauA} for Tau A, and Sect.~\ref{sec:pol_eff_moon} for the Moon.
The polarization is expected to be better constrained using the MFCI calibration, where the system operates under controlled conditions, free from systematics introduced by mapping strategies and reconstruction methods (BF for Tau A and AP for the Moon). However, on-sky measurements provide complementary characterization by capturing the instrument’s behavior under realistic observing conditions, including factors such as atmospheric loading, telescope optics, and scanning strategy, which are not fully replicated in ground-MFCI characterization.
The results for the ground and sky measurements are compatible at 1$\sigma$, while the results for the Moon present larger uncertainties. The primary issue with the Moon concerns evaluating the theoretical model, which accounts for changes in polarization emission due to the refraction index.

\subsection{Responsivity}
\label{sec:comparison_responsivity}

\begin{figure}
    \centering    
    \includegraphics[width=.95\columnwidth]{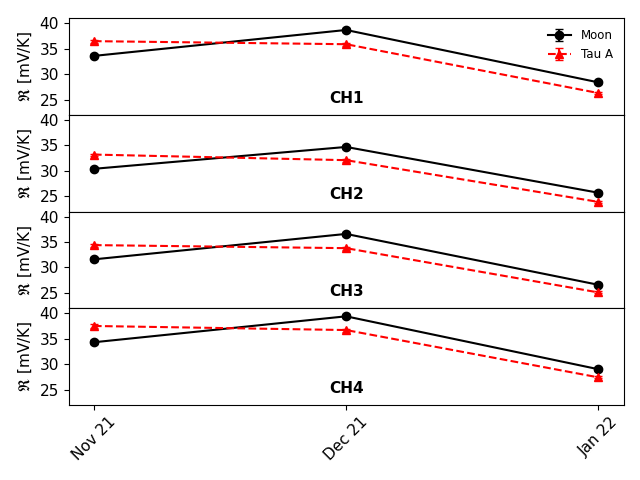}
    \caption{Evolution with time of the on-sky responsivity ($\mathfrak{R}$) calculated on Moon (black) and Tau A (red) observations. The results obtained on the four channels (CH) of TGI Pixel 23 are represented from top to bottom.}
    \label{fig:responsivity_vs_time}
\end{figure}

Figure~\ref{fig:responsivity_vs_time} presents the results on responsivity as a function of the observation period for each channel, including Tau A and the Moon.  
%/home/fasano/Dropbox/shared/projects/quijote/analysis/tfgi/responsivity_timeline/responsivity_vs_time.py
When comparing Tau A and the Moon, the responsivity presents discrepancies of $\sim$15\%. 
Overall, intercalibration within the channels is maintained, but the responsivity shows significant variations over time.
Furthermore, the results are coherent within the period and astronomical source, suggesting a common factor that alters the gain, such as the temperature of the backend.

The results are inconsistent between the astronomical sources and the ground measurements. This inconsistency arises from external factors that may alter the internal gain of the detectors, such as the surrounding temperature during observation. Specifically, we are not utilizing the complete observational setup for the ground tests, as the QT2 mirrors are not employed. Instead, we inject the diode directly into the focal plane, bypassing the QT2 optics. Nevertheless, we expect the optical efficiency to be very high due to the mirrors being highly under-illuminated.
There should be minimal loss due to spillover, and the beam is pseudo-Gaussian with sidelobes at <-25\,dB (\citealp{2017hsa9.conf...99R}). The large signal (600\,K) of the reference source may have introduced non-linear effects in the signal registration at the detector level.
Overall, intercalibration within the channels is maintained, but the responsivity shows significant variations over time. These systematics will be mitigated by introducing a reference calibration diode in the future.

\subsection{Polarization angle}

Table~\ref{tab:pol_angle} summarizes the results of the polarization angle characterization with the three different setups.
Section~\ref{sec:pol_ang_lab} details the results for ground characterization, Sect.~\ref{sec:pol_ang_crab} with Tau A, and Section~\ref{sec:pol_ang_moon} with the Moon.
We notice a deviation of up to $\sim$3\,deg in a three-month variability.
Finally, we expect a variation in the polarization angle across the bandwidth. Tau A, indeed, presents a synchrotron spectrum. At the same time, the diode may exhibit slight deviations from a perfect flat spectrum, primarily due to the directivity of the antenna to which it is coupled. The results from the different setups are compatible at $\sim$3$\sigma$. 
%This good agreement demonstrates how intercalibration within channels and the capabilities of each setup are very efficient.
Tau A provides the most reliable reference for the polarization angle because its emission is stable, well-characterized, and free from the geometric modeling uncertainties associated with lunar scattering. Therefore, the Tau A–based measurement offers the clearest on-sky calibration and ensures better control of systematics compared to the Moon.

\begin{table}[h]
    \caption{ Comparison of the results obtained with the three methods (MFCI ground calibration and on-sky calibration with Tau A and with the Moon) for the polarization efficiency (top) and the polarization angle (bottom). Different columns correspond to the different channels of the TGI Pixel 23.
    }
    \centering
    \small
    \begin{tabular}{rcccc}
        \hline \hline
        \multirow{2}{*}{} & \multicolumn{4}{c}{Polarization efficiency [\%]} \\
        & CH1 & CH2 & CH3 & CH4 \\
        \hline
        Ground & $96.2 \pm 0.3$ & $91.9 \pm 0.6$ & $96.7 \pm 0.3$ & $94.6 \pm 0.4$ \\
        Tau A & $91 \pm 4$ & $98 \pm 4$ & $97 \pm 6$ & $93 \pm 6$ \\
        Moon & 89 $\pm$ 16 & 88 $\pm$ 16 & 91 $\pm$ 16 & 86 $\pm$ 15 \\
        \hline
        \multirow{2}{*}{} & \multicolumn{4}{c}{Polarization angle [deg]} \\
        & CH1 & CH2 & CH3 & CH4 \\
        \hline
        Ground & $131.5 \pm 0.3$ & $-11.7 \pm 0.3$ & $42.9 \pm 0.3$ & $78.8 \pm 0.3$ \\
        Tau A & $132.0 \pm 1.4$ & $-15.1 \pm 1.3$ & $48.7 \pm 1.4$ & $73.9 \pm 1.5$ \\
        Moon& $127.2 \pm 2.2$ & $-15.5 \pm 2.5$ & $38.9 \pm 2.1$ & $75.1 \pm 1.7$ \\

        \hline
    \end{tabular}
    \label{tab:pol_angle}
\end{table}

\section{Conclusions}
\label{sec:conclusions}

This work presents the characterization of the TFGI experiment's polarization efficiency, responsivity, and polarization angle, focusing on the central detector of the TFGI, Pixel 23.
We begin with ground-based diode reference measurements to quantify polarization efficiency, responsivity, and polarization angle. Additionally, we analyze a dedicated dataset in which a rotating polarizer introduces multiple polarization angles to break degeneracies in the determination of instrumental angles. This dataset enables precise estimation of the detector’s polarization angle. We further use it to identify any systematic error angle associated with the phase switches relative to their expected configuration. Our results show that, within uncertainties, these error angles are consistent with 0\,deg at the $2\sigma$ level, and smaller than $\sim$2\,deg. Therefore, no additional correction is required for scientific exploitation of the data, ensuring polarization angle accuracy at the 1-deg level.

We extend our analysis using on-sky observations of Tau A and the Moon to further assess polarization efficiency, responsivity calibration, long-term responsivity stability, and polarization angle. We observe responsivity variations by up to a factor of $\sim$1.5 over a three-month period. Nevertheless, inter-channel responsivity ratios remain stable within each dataset, suggesting that a common system component may be driving these changes.

In particular, the Moon observations provide an independent check on the instrumental polarization calibration. By assuming a fixed polarization efficiency from Tau A, we estimate the effective refraction index of the Moon at 31\,GHz to be $n_{\rm{Moon}} = 1.209 \pm 0.007\, \text{(stat)}\,\pm0.009\,\text{(sys)}$, consistent with limb-weighted emission dominated by uppermost surface layer and previous measurement. This result aligns with previous findings that the apparent dielectric constant can vary significantly with surface roughness, reinforcing the Moon’s utility as a polarization calibrator at microwave frequencies.

This work highlights the need for implementing a real-time system to monitor responsivity in conjunction with calibrator observations.
This conclusion is motivated by the observed temporal variability of calibration parameters during commissioning, which demonstrates that continuous monitoring would be necessary to track and correct such effects in real time.
Following the example of the  MFI, we will install an internal diode calibrator to sample responsivity every 30 seconds. This system will provide relative responsivity calibration for tracking stability, while absolute calibration will continue to rely on astrophysical sources, such as Tau A. The new diode system will enhance the mitigation of systematics, yielding more robust polarization angle determination and improved temporal stability of the instrument’s response.

The scientific goal is to achieve an absolute calibration performance for TFGI comparable to that obtained with the QUIJOTE MFI 11\,GHz band \citep{mfiwidesurvey}, namely an internal calibration uncertainty of $\sim$1\% and an overall uncertainty of $\sim$5\% when including systematic errors associated with the modeling of the calibrators. For the polarization angle, we aim for an accuracy of about 1\,deg, similar to what is achieved at 11\,GHz with MFI. 
In this work, the absolute responsivity (derived independently from Tau A and from the Moon) is consistent at the few-percent level.
However, no internal calibration diode is currently installed, and therefore, the absolute responsivity is not yet stabilized across different observing periods.
The internal consistency of the responsivity calibration is presently limited to $<$8\% level due to residual systematic effects identified during commissioning. 
Regarding the polarization angle, we obtain uncertainties at the $\sim$1\,deg level, well within the $<$3\,deg requirement inferred from previous QUIJOTE analyses \citep{2013PhDT.......406L}.
These performances are adequate for the primary QUIJOTE science goals at these frequencies.

Finally, we note that the analyses presented here do not account for variations across the frequency band. A detailed characterization of the bandpass will be essential in future work, as unmodeled frequency dependence may bias the measured polarization angle and, consequently, the inferred angle errors.

\begin{acknowledgements}
Some of the results presented in this work are based on observations obtained with the QUIJOTE experiment (\url{doi.org/10.26698/quijote-mfi-dr1}). 
We thank the staff of the Teide Observatory for invaluable assistance in the commissioning and operation of QUIJOTE.
The QUIJOTE experiment is being developed by the Instituto de Astrofísica de Canarias (IAC),
the Instituto de Fisica de Cantabria (IFCA), and the Universities of Cantabria, Manchester and Cambridge.
Partial financial support was provided by the Spanish Ministry of Science and Innovation 
under the projects AYA2007-68058-C03-01, AYA2007-68058-C03-02,
AYA2010-21766-C03-01, AYA2010-21766-C03-02, AYA2014-60438-P,
ESP2015-70646-C2-1-R, AYA2017-84185-P, ESP2017-83921-C2-1-R,
PGC2018-101814-B-I00, PID2019-110610RB-C21, PID2020-120514GB-I00, 
IACA13-3E-2336, IACA15-BE-3707, EQC2018-004918-P, PID2023-150398NB-I00 and PID2023-151567NB-I00, the Severo Ochoa Programs SEV-2015-0548 and CEX2019-000920-S, the
Maria de Maeztu Program MDM-2017-0765, and by the Consolider-Ingenio project CSD2010-00064 (EPI: Exploring
the Physics of Inflation). We acknowledge support from the ACIISI, Consejeria de Economia, Conocimiento y 
Empleo del Gobierno de Canarias and the European Regional Development Fund (ERDF) under grant with reference ProID2020010108, and 
Red de Investigaci\'on RED2022-134715-T funded by MCIN/AEI/10.13039/501100011033.
This project has received funding from the European Union's Horizon 2020 research and innovation program under
grant agreement number 687312 (RADIOFOREGROUNDS), and the Horizon Europe research and innovation program under GA 101135036 (RadioForegroundsPlus).

This research made use of computing time available on the high-performance computing systems at the IAC. We thankfully acknowledge the technical expertise and assistance provided by the Spanish Supercomputing Network (Red Espa\~nola de Supercomputaci\'on), as well as the computer resources used: the Deimos/Diva Supercomputer, located at the IAC. This research used resources of the National Energy Research Scientific Computing Center, that is supported by the Office of Science of the U.S. Department of Energy under Contract No. DE-AC02-05CH11231. MFT acknowledges support from the Enigmass+ research federation (CNRS, Université Grenoble Alpes, Université Savoie Mont-Blanc). Some of the results in this work have been derived using the healpy and {\tt HEALPix} packages \citep{Healpix, Healpix2}. We have also used {\tt scipy} \citep{scipy}, {\tt emcee} \citep{emcee}, {\tt numpy} \citep{numpy}, {\tt matplotlib} \citep{matplotlib}, {\tt corner} \citep{corner} and {\tt astropy} \citep{astropy1,astropy2} \textsc{python} packages.
\end{acknowledgements}

\bibliographystyle{aa}
\bibliography{bibliography, quijote}

\appendix

\section{Effective refraction index of the Moon at 31\,GHz under smooth-surface assumption}
\label{sec:moon_refraction_index}

We estimate the effective refraction index of the Moon ($n_{\rm{Moon}}$) at 31\,GHz by exploiting its role as a polarization calibrator. Direct estimates of polarization efficiency from Moon observations present large uncertainties. To address this issue, we fix the polarization efficiency for each channel to the values derived from Tau A measurements (see Table~\ref{tab:TauA_211130}), which align with independent ground-based calibrations (see the top of Table~\ref{tab:pol_angle_total}), and solve for the value of $n_{\rm{Moon}}$ that minimizes the residuals between the observed and simulated Moon polarization maps.

The resulting estimates of the Moon’s refraction index are summarized in Table~\ref{tab:Moon:ni}.

\begin{table*}[ht]
    \caption{Effective refraction index of the Moon at 31\,GHz ($n_{\rm{Moon}}$).}
    \centering
    \small
    \begin{tabular}{rcccc}
\hline \hline
\multirow{2}{*}{Month}  &    \multicolumn{4}{c}{Moon refraction index @ 31\,GHz} \\
  & CH1     &    CH2     &    CH3     &    CH4     \\
\hline

Nov 21 & 1.211 $\pm$ 0.007 & 1.197 $\pm$ 0.006 & 1.204 $\pm$ 0.009 & 1.206 $\pm$ 0.010 \\
Dec 21 & 1.225 $\pm$ 0.007 & 1.208 $\pm$ 0.006 & 1.212 $\pm$ 0.010 & 1.210 $\pm$ 0.010 \\
Jan 22 & 1.215 $\pm$ 0.007 & 1.202 $\pm$ 0.006 & 1.209 $\pm$ 0.009 & 1.205 $\pm$ 0.010 \\

%Nov 21 & 1.187 $\pm$ 0.006 & 1.194 $\pm$ 0.006 & 1.180 $\pm$ 0.008 & 1.153 $\pm$ 0.007 \\
%Dec 21 & 1.200 $\pm$ 0.007 & 1.208 $\pm$ 0.006 & 1.185 $\pm$ 0.008 & 1.166 $\pm$ 0.008 \\
%Jan 22 & 1.191 $\pm$ 0.006 & 1.196 $\pm$ 0.006 & 1.180 $\pm$ 0.008 & 1.162 $\pm$ 0.007 \\ 

\hline
\end{tabular}

%$n_i$ - 17/09/22 & 1.061 $\pm$ 0.002 & 1.061 $\pm$ 0.002 & 1.056 $\pm$ 0.002 & 1.056 $\pm$ 0.002 \\
    \tablefoot{The refraction index is derived at multiple observation dates by fixing the polarization efficiency to the Tau A-derived values listed in Table~\ref{tab:TauA_211130}.}
    \label{tab:Moon:ni}
\end{table*}

From this procedure, we derive an effective refraction index of $n_{\rm{Moon}}=1.209 \pm 0.007\, \text{(stat)}\,\pm0.009\,\text{(sys)}$ To estimate the systematic uncertainty, we first reran the calculation varying the beam FWHM by 2\%, based on the TGI beam ellipticity analysis (see \citealp{mateo_phd}), obtaining a contribution of 0.005. In addition, we considered potential intensity-to-polarization leakage. Assuming a conservative worst-case scenario of 0.5\% leakage from intensity into both $Q$ and $U$, as expected during high-speed telescope scanning \citep{mateo_phd}, this effect contributes an additional 0.008 to the refraction index. These two contributions are combined in quadrature, yielding the quoted systematic uncertainty of 0.009.
Additional factors, such as the assumed polarization model, viewing geometry, surface heterogeneity, and limb dominance, which are not included in this analysis, may also contribute to systematic uncertainties. The quoted systematic uncertainty should therefore be regarded as a lower limit.

Interestingly, our estimated $n_{\rm{Moon}}$ aligns closely with the value of $n_{\rm{Moon}}\sim	1.25$ independently derived by \cite{mfipipeline} using observations of Tau A obtained with the Sardinia Radio Telescope (SRT). This consistency indicates a potential trend toward lower effective refraction indices when probing the Moon’s surface through polarization-sensitive measurements, particularly at microwave frequencies.

An earlier estimate by \cite{1967SvA....11..329L} reported a dielectric constant of $\epsilon_{\rm{Moon}} = 1.5 \pm 0.2$ at $\sim$38\,GHz, corresponding to $n_{\rm{Moon}} = \sqrt{\epsilon_{\rm{Moon}}} = 1.22 \pm 0.08$, based on a smooth, isotropic, non-magnetic lunar surface model. This result is compatible with ours within 1$\sigma$. Notably, the author emphasized that including diffuse scattering from small-scale surface roughness results in a higher effective dielectric constant. 

More recent studies generally report higher refraction indices, typically in the range $n_{\rm{Moon}} \sim 1.5$--$1.9$, based on laboratory measurements of Apollo regolith samples \citep[e.g.,][]{1973LPSC....4.3133O,1975E&PSL..24..394O,2012AdSpR..50.1607C}, detailed electromagnetic modeling \citep[e.g.,][]{1977NASSP.370..417S,1984Icar...60..568K}, and orbital observations from microwave radiometers aboard missions such as Chang’E-1/2 (see \citealp{2012Icar..219..194Z}). These works typically incorporate more complete dielectric models and consider the bulk properties of compacted regolith.

In contrast, our measurement is derived from polarized emission near the lunar limb. Consequently, our analysis probes the uppermost surface layer -- a porous, low-density region that is not well represented by the bulk permittivity of the deeper regolith. This interpretation is supported by spatial variations observed in microwave emissivity and dielectric constant maps \citep[e.g.,][]{2010ScChD..53.1365W}.

\section{Data visualization and structure}
\label{app:data_structure}
\begin{figure*}
    \centering
    \includegraphics[width=.8\textwidth]{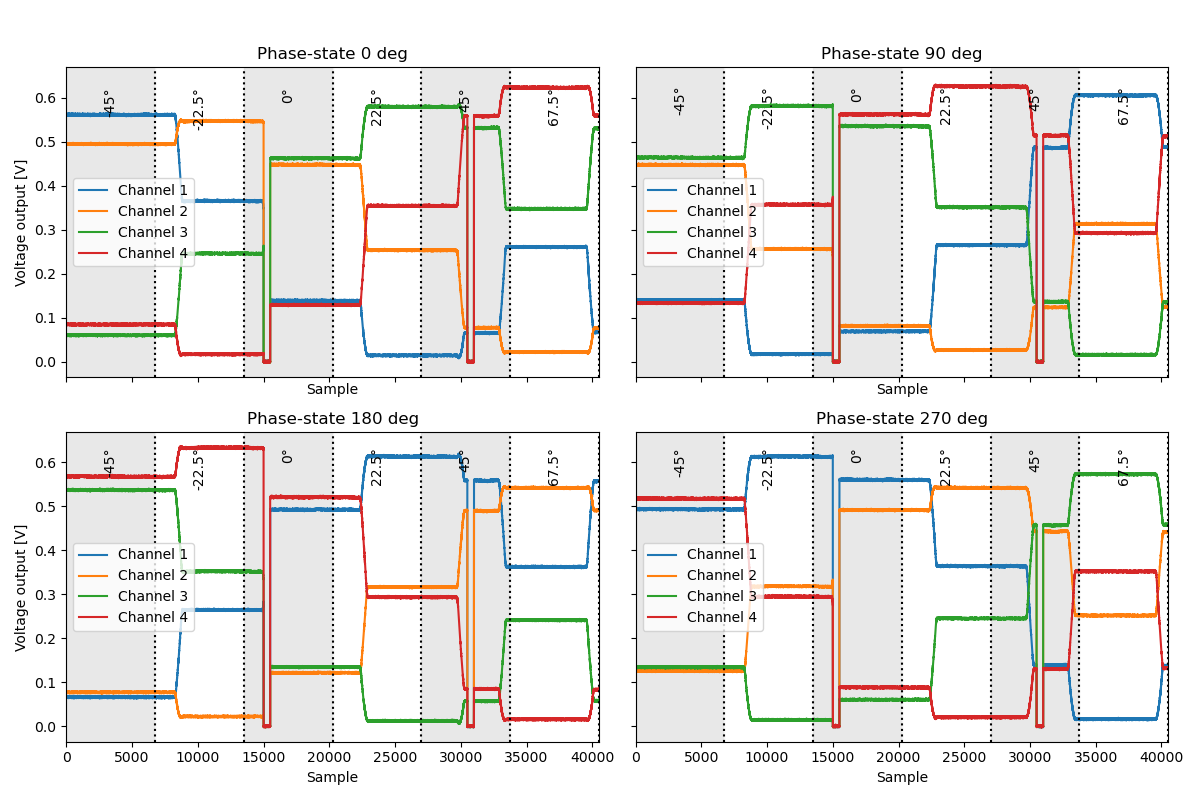}
    \caption{The data is displayed for each phase state, going from left to right, and from top to bottom: 0, 90, 180, and 270\,deg. Within each subplot, the output for the four channels is shown. The polarizer angle change is represented by alternating white and gray shades within each subplot.}
    \label{fig:overall_data}
\end{figure*}

In this Appendix, we clarify the data structure by visualizing the output data for each phase state and channel, segmented by the single polarizer angle within the shaded region. Figure~\ref{fig:overall_data} presents the results.   

\section{Stokes parameters equations and polarization efficiency}
\label{app:stokes}

We list all the equations for the Q and U Stokes parameters for each channel combination:

\label{app:eq_pol}
\begin{align}
    -Q_{\rm{TGI,FGI}} = V_{1,4}(90\,\rm{deg}) &- V_{1,4}(270\,\rm{deg}) \text{ ,} \\
    +U_{\rm{TGI,FGI}} = V_{1,4}(0\,\rm{deg}) &- V_{1,4}(180\,\rm{deg}) \text{ ,} \\
    +Q_{\rm{TGI,FGI}} = V_{2,3}(90\,\rm{deg}) &- V_{2,3}(270\,\rm{deg}) \text{ ,} \\
    -U_{\rm{TGI,FGI}} = V_{2,3}(0\,\rm{deg}) &- V_{2,3}(180\,\rm{deg}) \text{ ,} \\
    +Q_{\rm{TGI,FGI}} = V_{3,1}(0\,\rm{deg})  &- V_{3,1}(180\,\rm{deg}) \text{ ,} \\
    +U_{\rm{TGI,FGI}} = V_{3,1}(90\,\rm{deg}) &- V_{3,1}(270\,\rm{deg}) \text{ ,} \\
    -Q_{\rm{TGI,FGI}} = V_{4,2}(0\,\rm{deg}) &- V_{4,2}(180\,\rm{deg}) \text{ ,} \\
    -U_{\rm{TGI,FGI}} = V_{4,2}(90\,\rm{deg}) &- V_{4,2}(270\,\rm{deg}) 
\end{align}

\noindent where the subscript refers to the TGI and FGI channels, respectively.

Table~\ref{tab:all_pol_eff} shows the results of the polarization efficiency for each channel and polarizer angle. The missing data occurs when the polarization efficiency exceeds 100\% due to a spike in the first dataset at 45\,deg in the polarizer angle. Overall, the polarization efficiency remains well above the 90\% value, confirming a polarization loss across all combinations. The first columns present the results for the three datasets described in Sect.~\ref{sec:cal_lab}, while the last column summarizes the results by displaying the mean value (Mean CH) and standard deviation (STD) for each channel and polarizer angle across datasets.

\begin{table*}[]
\caption{Polarization efficiency ($\eta$) for each channel and polarizer angle from the datasets described in Sect.~\ref{sec:cal_lab}.  The mean value per channel is reported in the last column.}
\centering
\small
\begin{tabular}{ccccccc}
\hline \hline
CH & $\gamma$ [deg] & 
2022\_April\_06\_1 & 
2022\_April\_06\_2 &  
2022\_April\_07\_3 & 
Mean$\pm$STD\,[\%] &  Mean CH$\pm$STD\,[\%] \\ 
\hline
1  & -45   & 96.6 & 96.6 & 96.3 & 96.5$\pm$0.2 & \multirow{6}{*}{96.2$\pm$0.3} \\ \cline{1-6}
1  & -22.5 & 95.8 & 96.0 & 95.6 & 95.8$\pm$0.2 &  \\ \cline{1-6}
1  & 0     & 96.2 & 96.1 & 95.9 & 96.1$\pm$0.2 &  \\ \cline{1-6}
1  & 22.5  & 96.5 & 96.7 & 96.4 & 96.5$\pm$0.2 &  \\ \cline{1-6}
1  & 45    & -    & 96.6 & 96.2 & 96.4$\pm$0.3 &  \\ \cline{1-6}
1  & 67.5  & 96.1 & 96.1 & 95.8 & 96.0$\pm$0.2 &  \\ 
\hline \hline
2  & -45   & 92.2 & 92.2 & 91.8 & 92.1$\pm$0.2 & \multirow{6}{*}{91.9$\pm$0.6} \\ \cline{1-6}
2  & -22.5 & 92.3 & 92.4 & 92.0 & 92.2$\pm$0.2 &  \\ \cline{1-6}
2  & 0     & 91.7 & 91.7 & 91.3 & 91.6$\pm$0.2 &  \\ \cline{1-6}
2  & 22.5  & 91.0 & 91.3 & 90.8 & 91.0$\pm$0.2 &  \\ \cline{1-6}
2  & 45    & -    & 92.1 & 91.8 & 92.0$\pm$0.3 &  \\ \cline{1-6}
2  & 67.5  & 92.8 & 92.8 & 92.4 & 92.7$\pm$0.2 &  \\
\hline \hline
3  & -45   & 97.1 & 97.1 & 96.8 & 97.0$\pm$0.2 & \multirow{6}{*}{96.7$\pm$0.3} \\ \cline{1-6}
3  & -22.5 & 96.5 & 96.6 & 96.2 & 96.4$\pm$0.2 &  \\ \cline{1-6}
3  & 0     & 96.8 & 96.7 & 96.4 & 96.6$\pm$0.2 &  \\ \cline{1-6}
3  & 22.5  & 96.9 & 97.1 & 96.8 & 97.0$\pm$0.2 &  \\ \cline{1-6}
3  & 45    & -    & 96.9 & 96.5 & 96.7$\pm$0.3 &  \\ \cline{1-6}
3  & 67.5  & 96.3 & 96.4 & 96.0 & 96.2$\pm$0.2 &  \\ 
\hline \hline
4  & -45   & 94.8 & 94.8 & 94.5 & 94.7$\pm$0.2 & \multirow{6}{*}{94.6$\pm$0.4} \\ \cline{1-6}
4  & -22.5 & 95.1 & 95.3 & 94.8 & 95.1$\pm$0.2 &  \\ \cline{1-6}
4  & 0     & 94.6 & 94.6 & 94.2 & 94.5$\pm$0.3 &  \\ \cline{1-6}
4  & 22.5  & 93.8 & 94.0 & 93.6 & 93.8$\pm$0.2 &  \\ \cline{1-6}
4  & 45    & -    & 94.7 & 94.3 & 94.5$\pm$0.3 &  \\ \cline{1-6}
4  & 67.5  & 95.1 & 95.2 & 94.7 & 95.0$\pm$0.3 &  \\ 
\hline
\end{tabular}

\label{tab:all_pol_eff}
\end{table*}

\section{Polarization angle with diode reference}
\label{app:pol_ang_lab}

Table~\ref{tab:pol_ang} presents the value of the polarization angle for each channel and polarizer angle. The study details are provided in Sect.~\ref{sec:pol_ang_lab}. The study was conducted with the datasets 2022\_April\_06\_1, 2022\_April\_06\_2, and 2022\_April\_07\_3 described in Sect.~\ref{sec:cal_lab}.

\begin{table*}
\centering
\small
\caption{Polarization angle for each channel and polarizer angle from the datasets described in Sect.~\ref{sec:cal_lab}. The mean value per channel is reported in the last column. These values need to be corrected for the zero polarization angle ($\rm{pol\_ang_{TGI}}$).}
\begin{tabular}{ccccccc}
\hline \hline
CH & $\gamma$ [deg] & 
2022\_April\_06\_1 [deg] & 
2022\_April\_06\_2 [deg] &  
2022\_April\_07\_3 [deg] & 
Mean$\pm$STD [deg] & Mean CH$\pm$STD [deg]\\
\hline 
1 & -45.0 & 117.2 & 117.2 & 116.0 & 116.8 $\pm$ 0.6 & \multirow{6}{*}{116.8 $\pm$ 0.3} \\ \cline{1-6}
1 & -22.5 & 117.4 & 117.3 & 116.0 & 116.9 $\pm$ 0.7 & \\ \cline{1-6}
1 &   0.0 & 117.2 & 117.1 & 115.8 & 116.7 $\pm$ 0.8 & \\ \cline{1-6}
1 &  22.5 & 117.3 & 117.2 & 116.0 & 116.8 $\pm$ 0.7 & \\ \cline{1-6}
1 &  45.0 & -     & 117.4 & 116.1 & 116.8 $\pm$ 0.7 & \\ \cline{1-6}
1 &  67.5 & 117.4 & 117.4 & 116.1 & 116.6 $\pm$ 0.8 & \\ \hline \hline
2 & -45.0 & -26.2 & -26.2 & -27.4 & -26.6 $\pm$ 0.7 & \multirow{6}{*}{-26.4 $\pm$ 0.3} \\ \cline{1-6}
2 & -22.5 & -25.8 & -25.9 & -27.1 & -26.3 $\pm$ 0.7 & \\ \cline{1-6}
2 &   0.0 & -25.7 & -25.8 & -27.0 & -26.2 $\pm$ 0.8 & \\ \cline{1-6}
2 &  22.5 & -25.9 & -26.0 & -27.3 & -26.4 $\pm$ 0.8 & \\ \cline{1-6}
2 &  45.0 & -     & -26.2 & -27.4 & -26.8 $\pm$ 0.6 & \\ \cline{1-6}
2 &  67.5 & -25.9 & -25.8 & -27.2 & -26.3 $\pm$ 0.8 & \\ \hline \hline
3 & -45.0 & 27.7  & 27.7 & 26.4 & 27.2 $\pm$ 0.7 & \multirow{6}{*}{27.4 $\pm$ 0.3} \\ \cline{1-6}
3 & -22.5 & 27.9  & 27.8 & 26.6 & 27.4 $\pm$ 0.7 & \\ \cline{1-6}
3 &   0.0 & 27.8  & 27.7 & 26.4 & 27.3 $\pm$ 0.8 & \\ \cline{1-6}
3 &  22.5 & 27.9  & 27.8 & 26.5 & 27.4 $\pm$ 0.8 & \\ \cline{1-6}
3 &  45.0 & -     & 27.9 & 26.7 & 27.3 $\pm$ 0.6 & \\ \cline{1-6}
3 &  67.5 & 27.9  & 28.0 & 26.6 & 27.5 $\pm$ 0.8 & \\ \hline \hline
4 & -45.0 & 64.2  & 64.2 & 63.0 & 63.8 $\pm$ 0.7 & \multirow{6}{*}{64.1 $\pm$ 0.3} \\ \cline{1-6}
4 & -22.5 & 64.7  & 64.6 & 63.4 & 64.2 $\pm$ 0.7 & \\ \cline{1-6}
4 &   0.0 & 64.9  & 64.8 & 63.5 & 64.4 $\pm$ 0.7 & \\ \cline{1-6}
4 &  22.5 & 64.7  & 64.6 & 63.4 & 64.2 $\pm$ 0.8 & \\ \cline{1-6}
4 &  45.0 & -     & 64.5 & 63.2 & 63.9 $\pm$ 0.7 & \\ \cline{1-6}
4 &  67.5 & 64.7  & 64.7 & 63.4 & 64.3 $\pm$ 0.8 & \\ \hline
\end{tabular}
%\tablefoot{The uncertainties are calculated as standard deviations (STDs).}
\label{tab:pol_ang}
\end{table*}

\section{Phase error fitting visualization and comparison with the data}
\label{app:fitting_model}
\begin{figure*}[ht]
        \centering
        \includegraphics[width=.75\textwidth]{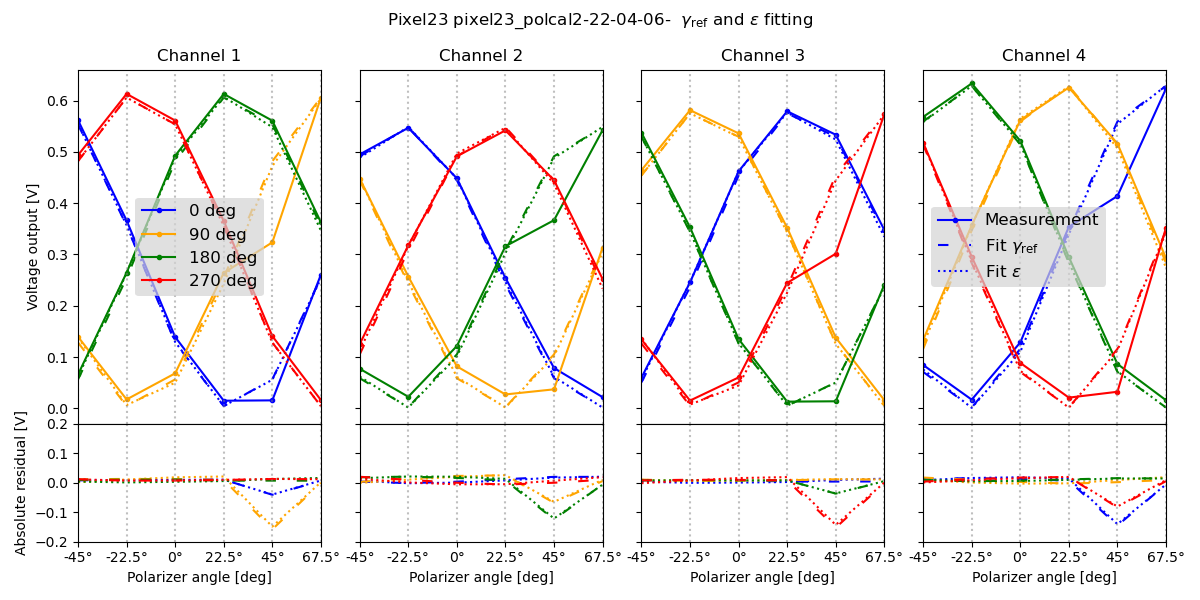}
        \caption{Visual representation of the fitting process for dataset Pixel23\_2022\_April\_06\_1. The recorded data are shown with continuous lines, while the fitting model incorporating $\gamma_{\text{ref}}$ is illustrated with dashed lines, and the model integrating $\epsilon$ is displayed with dotted lines. We do not use this dataset for calculating $\epsilon$ due to the presence of spikes.}
        \label{fig:fitting_feedback1}
\end{figure*}

\begin{figure*}[ht]
        \centering
        \includegraphics[width=.75\textwidth]{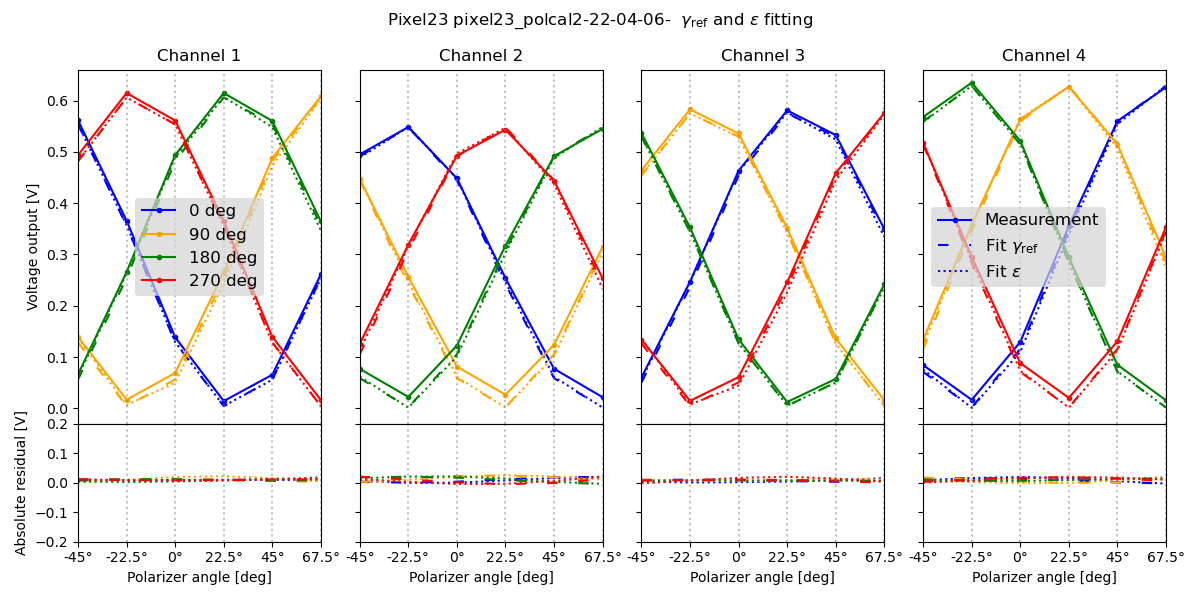}
        \caption{ Same as Fig.~\ref{fig:fitting_feedback1} but for dataset Pixel23\_2022\_April\_06\_2.}
        \label{fig:fitting_feedback2}
\end{figure*}

\begin{figure*}[ht]
        \centering
        \includegraphics[width=.75\textwidth]{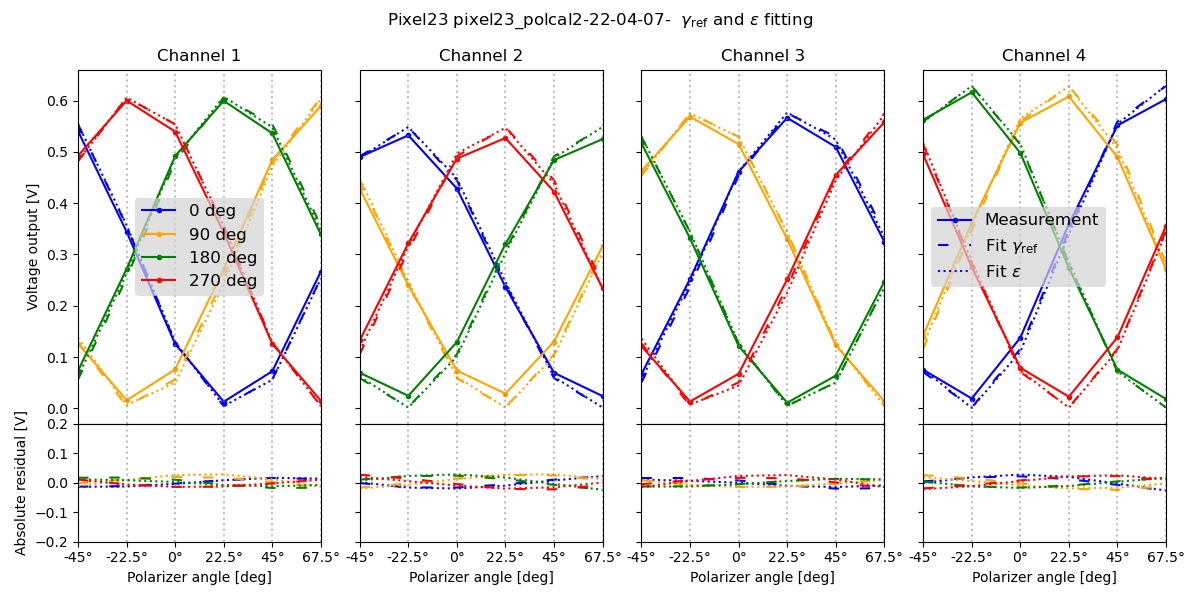}
        \caption{ Same as Fig.~\ref{fig:fitting_feedback1} and Fig.~\ref{fig:fitting_feedback2} but for dataset Pixel23\_2022\_April\_07\_3.}
        \label{fig:fitting_feedback3}
\end{figure*}

This section presents the visual results of the fitting function applied to the actual data in Fig.~\ref{fig:fitting_feedback1},\ref{fig:fitting_feedback2},\ref{fig:fitting_feedback3}. For the datasets 2022\_April\_06\_1, 2022\_April\_06\_2, and 2022\_April\_07\_3 described in Sect.~\ref{sec:cal_lab}.
\newpage

\begin{figure}[ht]
   \centering
   \begin{subfigure}[b]{0.49\textwidth}
       \centering
       \includegraphics[width=\textwidth]{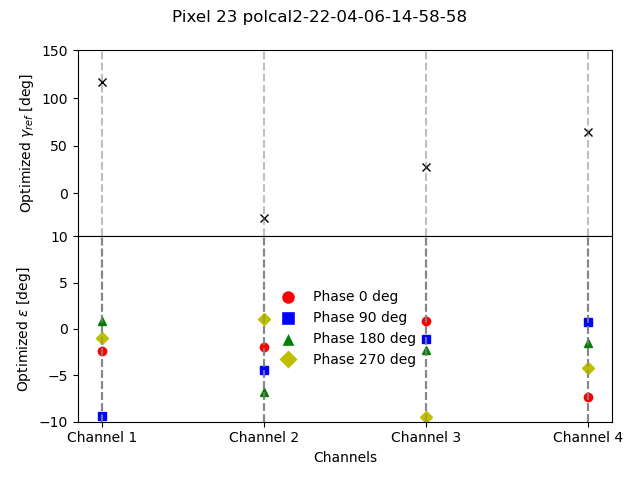}
       \caption{Pixel23\_2022\_April\_06\_1}
   \end{subfigure}
   \begin{subfigure}[b]{0.49\textwidth}
       \centering
       \includegraphics[width=\textwidth]{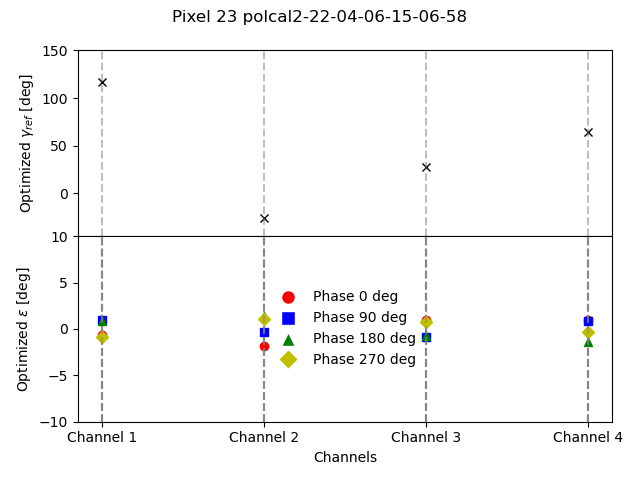}
       \caption{Pixel23\_2022\_April\_06\_2}
   \end{subfigure}
   \begin{subfigure}[b]{0.49\textwidth}
       \centering
       \includegraphics[width=\textwidth]{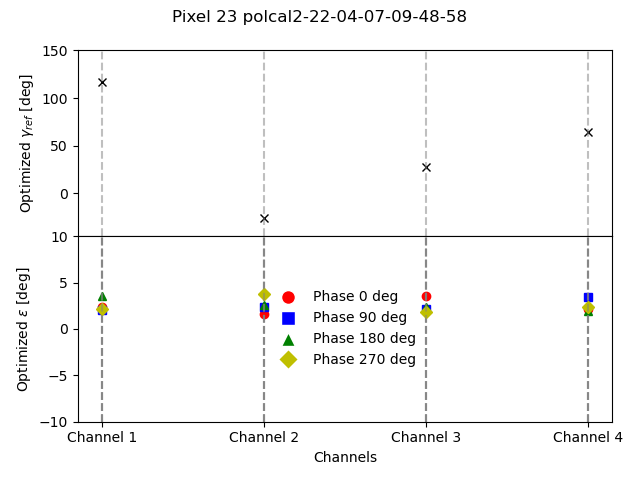}
       \caption{Pixel23\_2022\_April\_07\_3}
   \end{subfigure}
   \caption{The average error angle correction, $\epsilon$, in the three datasets along with their standard deviations for each channel. The average $\gamma_{\rm{ref}}$ reported here ($<\gamma_{\rm{ref}}>$) has to be corrected to obtain the on-sky polarization angle, $\phi_{\rm{c}}$ (see Eq.~\eqref{eq:gamma_ref}). The first dataset is not valid for the $\epsilon$ calculation due to the occurrence of a spike in the polarizer angle at 45\,deg.}
   \label{fig:epsilon_datasets}
\end{figure}

Including the $\epsilon$ parameters in the fitting does not significantly affect the quality of the results regarding residuals, based on an equivalent 0-deg angle correction. We observe inconsistencies in the fitting process as we approach the 0-V signal. This effect arises because the polarization efficiency does not reach 100\% (as reported in Appendix~\ref{app:eq_pol}).
The results of these fittings are displayed in Figure \ref{fig:epsilon_datasets}.

\end{document}